\documentclass[preprint,amsmath,amssymb,aps]{revtex4-1}

\usepackage{changepage}
\usepackage{dcolumn}
\usepackage[table]{xcolor}
\usepackage{pgfplots}
\usepackage{floatrow}
\usepackage{floatrow}   
\usepackage{subcaption}
\usepackage{graphicx,xcolor} 

\usepackage[margin=1in]{geometry}
\usepackage{amsmath}
\usepackage{amssymb}
\usepackage{pgf}
\usepackage[latin1]{inputenc}
\usepackage{verbatim}
\usepackage{physics}
\usepackage{MnSymbol}
\usepackage{wasysym}
\usepackage{cleveref}
\usepackage{ragged2e}
 
\usetikzlibrary{positioning, decorations.pathmorphing}
\usetikzlibrary{shapes,arrows,shadows,positioning}
\usetikzlibrary{backgrounds}
\usetikzlibrary{external}
\definecolor{mynavy}{HTML}{000080}
\definecolor{darkred}{HTML}{8B0000}
\definecolor{mygreen}{HTML}{006400}
\definecolor{mygold}{HTML}{B8860B}

\makeatletter
\newcommand\footnoteref[1]{\protected@xdef\@thefnmark{\ref{#1}}\@footnotemark}
\makeatother

\makeatletter
\renewcommand\@make@capt@title[2]{%
\@ifx@empty\float@link{\@firstofone}{\expandafter\href\expandafter{\float@link}}%
{\textbf{#1}}\@caption@fignum@sep#2\quad}%
\makeatother

\newtheorem{proposition}{Proposition}

\newcolumntype{d}[1]{D..{#1}}

\begin{document}

\preprint{APS/123-QED}

\title{Cooperation dynamics of generalized reciprocity in state-based social dilemmas}

\author{Viktor Stojkoski$^{1}$}
\author{Zoran Utkovski$^{2}$}
\author{Lasko Basnarkov$^{1,3}$}
\author{Ljupco Kocarev$^{1,3}$}

\affiliation{
$^{1}$Macedonian Academy of Sciences and Arts, P.O. Box 428, 1000 Skopje, Republic of Macedonia}%
\affiliation{
$^{2}$Fraunhofer Heinrich Hertz Institute, Einsteinufer 37, 10587, Berlin, Germany}
\affiliation{
$^{3}$Faculty of Computer Science and Engineering, Ss. Cyril and Methodius University, P.O. Box 393, 1000 Skopje, Republic of Macedonia\\}%

\date{\today}

\begin{abstract}
 We introduce a framework for studying social dilemmas in networked societies where individuals follow a simple state-based behavioral mechanism based on generalized reciprocity, which is rooted in the principle ``help anyone if helped by someone''. Within this general framework, which applies to a wide range of social dilemmas including, among others, public goods, donation and snowdrift games, we study the cooperation dynamics on a variety of complex network examples. By interpreting the studied model through the lenses of nonlinear dynamical systems, we show that cooperation through generalized reciprocity always emerges as the unique attractor in which the overall level of cooperation is maximized, while simultaneously exploitation of the participating individuals is prevented. The analysis elucidates the role of the network structure, here captured by a local centrality measure which uniquely quantifies the propensity of the network structure to cooperation by dictating the degree of cooperation displayed both at microscopic and macroscopic level. We demonstrate the applicability of the analysis on a practical example by considering an interaction structure that couples a donation process with a public goods game.
\end{abstract}

\maketitle

\section{Introduction}

 A \textit{social dilemma} arises in a situation where individual decisions are at odds with the performance of the collective. As such it represents the epitome for studying the emergence and stability of cooperative behavior in both natural and artificial systems~\cite{capraro2013model,wang2014different,szabo2018social}.

Ever since the publication of Darwin's epochal work~\cite{darwin1888descent}, the appearing paradox of cooperation in social dilemmas has been at the focus of the research community. The insight that all major transitions in biological evolution, from simple to complex structures, are characterized by some degree of cooperation and sacrifice~\cite{Smith-1973}, has subsequently led to major advances in the field. However, despite decades of investigation the cooperation paradigm is still regarded as one of the most challenging issues currently faced by scientists~\cite{Pennisi-2005}.

 A particular setting whose study is of great value to the complex systems community are situations where an individual has repeated encounters within not necessarily the same group or interaction structures. In this context, the concepts of direct reciprocity (``help someone who has helped you before'')~\cite{Trivers-1971} and indirect reciprocity (``help someone who is helpful'')~\cite{nowak2005evolution} have been able to provide solutions for the emergence of cooperation in essentially disparate types of social dilemmas. Although initially structured for encounters resembling a prisoner's dilemma~\cite{Axelrod-1981}, both mechanisms have been extended to account for a disperse class of interaction structures that are ubiquitous in natural systems (see Refs.~\cite{van2012emergence,hauert2006synergy,killingback2002continuous,yoeli2013powering,hilbe2014cooperation,hilbe2015evolutionary,pan2015zero}). 

While being of significant theoretical value, the extent to which direct and indirect reciprocity are able to explain cooperation in real-life systems has recently been put into question~\cite{vanDoorn-2012,Taborsky-2016}. The reason for this is that the application of the rules is costly (in terms of memory and processing requirements), in the sense that they demand high cognitive abilities such as recognition of the group with which an individual is engaged in reciprocal mechanisms or knowledge about the interaction outcomes. This, for example, limits the emergence of cooperation in systems where there is randomness in the interactions and the individuals do not posses the cognitive prowess to acknowledge with whom they play~\cite{Taborsky-2016}.

\begin{figure}[t]
\includegraphics[width=7.7cm]{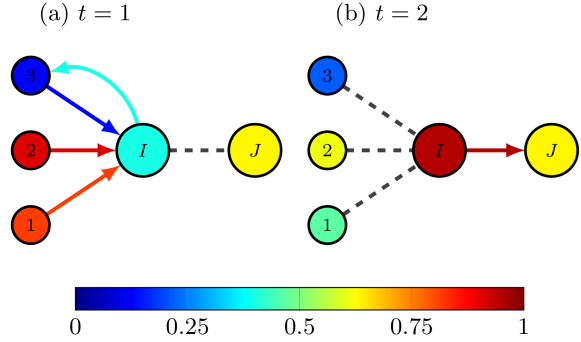}
\caption{The concept of generalized reciprocity explained by the (random) interactions within a group of 5 individuals. Solid directed lines are active relationships, whereas dashed are inactive in the respective round. In each round the individuals are colored according to their willingness for cooperation described by the internal state. As shown, in round $t=1$ individual $I$ exchanges cooperative experiences with $1$, $2$ and $3$. This results in changes in the internal states of the interacting individuals and the level of cooperation that is provided when tasked in $t = 2$. }
\label{fig:gen-rep-explanation}
\end{figure}

To tackle this problem, the concept of \textit{generalized reciprocity}, formally defined as the rule of ``help anyone if helped by someone'', has been developed~\cite{pfeiffer2005evolution}. This concept, whose roots lie within ``upstream reciprocity''~\cite{Boyd-1989,Nowak-2005} has been fully explored in Refs.~\cite{vanDoorn-2012,Taborsky-2016}. The intrinsic feature that may favor generalized reciprocity over others is that the proximate mechanism behind it may be explained by the changes of an individuals' physiological condition~\cite{Rutte-2007,Bartlett-2006,isen1987positive}. In other words, the decision of an individual whether to cooperate or not, is based on an \textit{internal cooperative state} which captures its past experience. To shed light on the dynamical process behind state-based generalized reciprocity, we provide an intuitive illustration in Fig.~\ref{fig:gen-rep-explanation}. 

This simple behavioral mechanism may apply to a wide range of dynamical interaction structures. Specifically, the internal state may mimic aggregated fitness in biological systems, wealth or well-being in animal and human societies, or energy level in artificial (e.g. communication) systems. These observations are complemented with convincing empirical evidence suggesting the possibility that such a mechanism may have indeed emerged in natural systems by evolution~\cite{Rutte-2007,Leimgruber-2014,Gfrerer-2017,Bartlett-2006,Stanca-2009}. 

Nevertheless, the development of theoretical models has been lacking and so far only pairwise linear interactions between individuals~\cite{Barta-2011,Utkovski-2017} have been considered. Other ubiquitous scenarios that describe a social dilemma as group interactions~\cite{Perc-2017}, the possibility of nonlinear payoffs~\cite{killingback2002continuous,hauert2006synergy} or even intertwined combinations of them~\cite{milinski2012interaction}, have been eschewed.

\subsection{Our contribution}

Motivated by the early exploits of Motro~\cite{motro1991co} on public goods, and the insights provided in~\cite{kerr2004altruism,hilbe2014cooperation,hilbe2015evolutionary}, here we revisit the concept of social dilemmas within a society that follows a state-based behavioral mechanism in the spirit of~\cite{Utkovski-2017}. We implement this framework to examine the applicability of the behavioral mechanism in a wide class of payoff structures, and thus study the general setup that is required for state-based generalized reciprocity to promote cooperation. Under this rather general model, the well-studied problems of prisoner's dilemma~\cite{Nowak-2006five}, public goods game~\cite{Perc-2017} and the common-pool resource problem~\cite{ostrom2015governing} arise as special cases (see Appendix~A for simple examples). Moreover, the model can easily accommodate different combinations of interaction structures and/or payoff functions and capture their dynamics. Even though such structures have been widespread in natural systems~\cite{killingback2002continuous}, the role of generalized reciprocity in their cooperation dynamics is yet to be explored. 

By treating the problem from the perspective of dynamical systems, here we develop a systematic way to investigate the implications created by these structures. Our model reveals that, under some simple assumptions, cooperation through generalized reciprocity emerges as the unique attractor that maximizes the global level of cooperation while at the same time preventing each participating individual from being exploited by the environment. The analysis elucidates the role of the interaction topology, here described via a \textit{complex network}, which covers the standardly studied ``mixed population'' and ``regular lattice'' models as special cases. It turns out that here a critical role plays a specific \textit{network centrality measure} which quantifies the propensity of the network structure to cooperation, by dictating the degree of cooperation displayed both at microscopic and macroscopic level. 

The structure of this paper is as follows. In Sec.~II we introduce the interaction structure, the state-based social dilemma and the behavioral update rule. In Sec.~III we provide an analytical study of the asymptotic behavior of the cooperation dynamics in a population consisting of both individuals that follow a state-based update rule and unconditional defectors. We examine two phase transitions, extinction of cooperation and unconditional cooperation, which show to critically capture the properties of the behavioral mechanism based on generalized reciprocity. In Sec.~IV we address in detail a practical example by considering an interaction structure that couples a donation process with a public goods game played on  random graphs. The last section summarizes our findings. Further examples of social dilemmas with some additional analysis and theoretical background are provided in Appendices~A-C. 

\section{Model}

\subsection{Interaction structure} 

We consider a class of dynamical models constituted of a finite population $\mathcal{N}$ of $N$ individuals. The models run in discrete time and explain the evolution of the $N$ dimensional vector $\mathbf{p}(t) \in \left[ 0,1 \right]^N$, where the $i$-th entry, $\mathrm{p}_i(t)$, describes the internal state of individual $i$ in round $t$. According to our representation the individual payoff generated in each time step, is a function of the population state vector, 
\begin{align}
\mathrm{y}_{i,\mathcal{G}} \left( t \right) &= b_{i,\mathcal{G}} \left( \mathbf{p}(t) \right) - c_{i,\mathcal{G}} \left( \mathrm{p}_i(t) \right),
\label{eq:payoff-general}
\end{align}
where $b_{i,\mathcal{G}}$ and $c_{i,\mathcal{G}}$ are respectively the benefit and cost function of individual $i$, both parametrized by a connected graph $\mathcal{G} \left( \mathcal{N}, \mathcal{E} \right)$. The graph is defined by a set of vertices $\mathcal{N}$, corresponding to the set of individuals, and $\mathcal{E} \subseteq \mathcal{N} \times \mathcal{N}$ is the set of edges which determines the pairwise relationships between individuals.

The generality of the model is captured by the freedom in the choice of the benefit function $b_{i,\mathcal{G}}$ and the cost function $c_{i,\mathcal{G}}$. The only constraint that we thereby make is that we restrict both functions to be sufficiently smooth and adhere to several simple assumptions that define a social dilemma. In particular, we assume that the benefit function is non-decreasing with respect to each coordinate-wise projection that is not $i$, with the note that it is strictly increasing for some projections that are defined through the network topology and the rules of interaction. This implies that $i$ gains with the increase in the willingness for cooperation of a particular group of other individuals. We name this group the $l$-neighborhood of $i$ and represent it as $\mathcal{N}^{(l)}_i$. Moreover, we restrict the cost function to be an increasing function of $\mathrm{p}_i(t)$ in order to capture the social setting of paying a higher cost for the increments in the benefits of others. Finally, both functions should satisfy $b_{i,\mathcal{G}} \left( \mathbf{0} \right) = c_{i,\mathcal{G}} \left( 0 \right) = 0$ so as indicating that nothing happens when everyone defects unconditionally.

In one possible physical interpretation, the model~(\ref{eq:payoff-general}) describes the payoff of a continuous game with deterministic (i.e fixed) interactions between the individuals at each time step and where each individual has a continuum of behavioral strategies to choose from. This is, for example, the case with the continuous iterative prisoner's dilemma. In another interpretation, (\ref{eq:payoff-general}) provides a deterministic approximation for the steady state of stochastic interactions among the individuals, with random payoffs generated by the individual internal states. In this context, $b_{i,\mathcal{G}}$ and $c_{i,\mathcal{G}}$ are affine maps with respect to the random variables. The second interpretation, that we will closely follow, is aligned to the concept of generalized reciprocity, in the sense that it describes the individual payoffs as a function of the benefit they obtain through random interactions with individuals from their neighborhood, without explicit records to the contributions during these interactions.

\subsection{The social dilemma}

The definitions of the benefit and cost functions offer a realistic representation for a plethora of real-life situations. However, they alone do not describe the social dilemma. To set the stage for this phenomena, we must define two additional conditions that have to be present within the payoff structure.

First, for every individual there should exist a point beyond which cooperation is costly, i.e. beyond this point the individual is better-off by not increasing the willingness to cooperate, while expecting cooperation from others. This implies that, full cooperation by all individuals does not belong to the set of Nash equilibria, defined as the set of points that maximize the individuals' payoff given the set of available actions of all other individuals. Formally, we define this condition as the point $\mathbf{p} \in \left[ 0,1 \right]^N $, such that for every other point $\mathbf{\hat{p}}$ satisfying $\mathrm{proj}_j \left(\mathbf{\hat{p}} \right) \geq \mathrm{proj}_j \left(\mathbf{p} \right)$ for all $j \in \mathcal{N}^{(l)}_i$, where $\mathrm{proj}_j \left(\mathbf{\hat{p}} \right)$ is the $j$-th coordinate-wise projection of $\mathbf{\hat{p}}$ , the rate at which the benefit of individual $i$ changes is lower than the change in the cost when only its internal state is slightly perturbed while everything else is kept constant. In other words, for all those $\mathbf{\hat{p}}$,
\begin{align}
\pdv{b_{i,\mathcal{G}} (\mathbf{\hat{p}})}{\mathrm{p}_i} < \dv{c_{i,\mathcal{G}}(\mathrm{\hat{p_i}})}{\mathrm{p}_i}.
\label{eq:dilemma_condition_1}
\end{align}

Second, for all $\mathbf{\hat{p}} \in \left[ 0,1 \right]^N $, the sum over the changes in benefits of all $j \in \mathcal{N}$ must be greater than or equal to the change in cost provided by $i$,
\begin{align}
\sum_j \pdv{b_{j,\mathcal{G}} (\mathbf{\hat{p}})}{\mathrm{p}_i}  \geq \dv{c_{i,\mathcal{G}} (\mathrm{\hat{p_i}})}{\mathrm{p}_i},
\label{eq:dilemma_condition_2}
\end{align}
with the strict inequality holding when $\mathbf{\hat{p}} = \mathbf{0}$.

Condition~(\ref{eq:dilemma_condition_2}) together with the definitions of the benefit and cost functions, $b_{i,\mathcal{G}}$ and $c_{i,\mathcal{G}}$, reveals that full cooperation by all individuals, $\mathrm{p}_i(t) = 1$, for all $i$ and $t$, is an efficient solution, i.e. it is a solution that maximizes the overall population payoff. Considered together these two conditions build a structure under which egoistic behavior leads to depletion of the performance of the overall system, which is exactly the social dilemma metaphor.

A typical example for an interaction mechanism that is easily captured by this representation is the snowdrift game, where cooperation is a favorable trait for an individual as long as the number of cooperative neighbors is low~\cite{Santos-2005}. Another group of mechanisms that are modeled when condition (\ref{eq:dilemma_condition_1}) holds for all $\mathbf{\hat{p}} \in \left[ 0,1 \right]^N$ and condition (\ref{eq:dilemma_condition_2}) is a strict inequality for every $\mathbf{\hat{p}}$ are the prisoner's dilemma~\cite{Santos-2005}, the common pool resource problem~\cite{ostrom2015governing}, the public goods game~\cite{Perc-2017}, as well as extensions and combinations of these three interactions structures that constrain the cost of an individual to be a function only of its own state are covered (see for example~\cite{milinski2012interaction,motro1991co,hilbe2014cooperation,hilbe2015evolutionary,Doebeli-2005,archetti2012game,battiston2017determinants}). In Appendix~A we provide a short overview for state-based social dilemmas that can be described through the first three interaction structures, while a comprehensive example for the public goods game will be analyzed in Sec.~IV.

\subsection{Behavioral update mechanism}

In our scenario, the individuals follow the state-based behavioral update introduced in~\cite{Utkovski-2017}, apart for a fraction of the population (a set $\mathcal{D}$) of unconditional defectors (with $\mathrm{p}_i (t) = 0$, $i\in \mathcal{D}$, for all $t$). 

The behavioral update rule describes the cooperative state of individual $i$ at time $t+1$ as a function of its' accumulated payoff $\mathrm{Y}_{i,\mathcal{G}}(t)$ by time~$t$ 
\begin{align}
\mathrm{p}_i(t+1)=f_i \left(\mathrm{Y}_{i,\mathcal{G}}(t)\right),
\label{eq:update}
\end{align}
where $\mathrm{Y}_{i,\mathcal{G}}(t)=\mathrm{Y}_{i,\mathcal{G}}(t-1)+\mathrm{y}_{i,\mathcal{G}}(t)$, with $\mathrm{Y}_{i,\mathcal{G}}(0)$ being the initial condition and $\mathrm{y}_{i,\mathcal{G}}(0)=0$. 

We assume that the function $f_i :\mathbb{R} \to \left(0, 1 \right)$, which maps the payoff $\mathrm{Y}$ to the willingness for cooperation $\mathrm{p}_i$, is continuous on the interval $\left(0, 1 \right)$ and has a continuous inverse (i.e. is a homeomorphism). Additionally, it is monotonically increasing with $\lim_{\mathrm{Y}\to -\infty} f_i(\mathrm{Y}) = 0$ and $\lim_{\mathrm{Y}\to \infty} f_i(\mathrm{Y}) = 1$. 

An example of a function with the above properties (which, for example, is often used for modeling in biology and ecology) is the logistic function $f_i(\mathrm{Y})=\left[1+e^{-\kappa_i(\mathrm{Y}-\omega_i)}\right]^{-1}$, where the parameters $\kappa_i$ and $\omega_i$ define the steepness, respectively the midpoint of the function.

This rule provides a simple description for the cooperative behavior in a wide range of dynamical interaction structures. Specifically, the internal state may mimic aggregated fitness in biological systems, wealth or well-being in animal and human societies, or energy level in artificial systems~\cite{Utkovski-2017}. Its advantage lies in its simplicity since (\ref{eq:update}) can easily be described as a Markovian process where an individual only has to know its present state and payoff in order to determine the next action. This is significantly different from other reciprocal update rules. For instance, in certain interaction structures direct reciprocity requires extensive memory requirements to record own and opponents' actions in order for cooperation to thrive~\cite{hilbe2017memory}.

\section{Results}
\subsection{Cooperation dynamics}

We begin the analysis by studying the properties of the steady state solution $\mathbf{p}^*$. We will thereby assume that the interaction structure is non-degenerate, which, in our terms, means that the Jacobian of the individual payoff functions accounting only for the individuals that follow the generalized reciprocity update rule, $\mathbf{J_y}^{\setminus\mathcal{D}}(\mathbf{p})$, at the point $\mathbf{0}$ is nonsingular. In fact, it is easy to notice that as a consequence of~(\ref{eq:dilemma_condition_2}), this assumption always holds when the whole population follows~(\ref{eq:update}).

Now, in steady state, for the individuals adhering to (\ref{eq:update}), i.e. the non-defectors, it holds 
\begin{align}
\mathrm{p}^*_i&=f_i\left(f_i^{-1}(\mathrm{p}^*_i)+b_{i,\mathcal{G}} \left( \mathbf{p}^* \right) - c_{i,\mathcal{G}} ( \mathrm{p}^*_i)\right)
\label{eq:iterative_steady_state}.
\end{align}
By applying the inverse map, we obtain
\begin{align}
f_i^{-1}\left(\mathrm{p}_i^*\right)&=f_i^{-1}\left(\mathrm{p}_i^*\right)+b_{i,\mathcal{G}} \left( \mathbf{p}^* \right) - c_{i,\mathcal{G}} \left( \mathrm{p}^*_i \right).
\label{eq:steady_state_individual_inverse}
\end{align}
For the above equation to hold, it is required that 1)~$\mathrm{y}^*_{i,\mathcal{G}}\doteq b_{i,\mathcal{G}} \left( \mathbf{p}^* \right) - c_{i,\mathcal{G}} \left( \mathrm{p}^*_i \right)=0$, unless either 2)~$\mathrm{p}_i^*=1$ (i.e. $\mathbf{\mathrm{Y}}_{i,\mathcal{G}}^*=\infty$), or 3)~$\mathrm{p}_i^*=0$ (i.e.$\mathbf{\mathrm{Y}}_{i,\mathcal{G}}^*=-\infty$). 

 It is easy to verify that condition 3) is a special case of condition 1) when $\mathrm{p}^*_j=0$ for all $j \in \mathcal{N}^{(l)}_i$. Indeed, when $\mathrm{p}^*_i=0$, it must be that $y^*_{i,\mathcal{G}}\leq 0$. By the definition of (\ref{eq:update}), the condition $y^*_{i,\mathcal{G}} < 0$ implies $\mathrm{p}^*_i=0$ which, on the other hand implies $y^*_{i,\mathcal{G}} \geq 0$, leading to a contradiction. Therefore, whenever $\mathrm{p}^*_i=0$ it must hold that $ y^*_{i,\mathcal{G}} = 0$, which is true if and only if $\mathrm{p}^*_j=0$ for all $j \in \mathcal{N}^{(l)}_i$.

These conditions, together with the (strict) monotonicity of the function $f_i$ and the assumption that the interactions are non-degenerate, reveal that each individual $i$ conforming to (\ref{eq:update}) will increase, respectively decrease, its willingness for cooperation (based on its environment, i.e. on the state dynamics of other individuals), until the system reaches a steady state where it is either satisfied $\mathrm{y}^*_{i,\mathcal{G}} = 0$ or $\mathrm{p}^*_{i} = 1$, for all $i \not\in \mathcal{D}$. 
In other words, each individual following the generalized reciprocity rule will cooperate with the maximal willingness, given its environment, so as it is not exploited. In this context, the steady state $\mathbf{p}^*$ may be interpreted as a solution of the constrained optimization problem of maximizing the global level of cooperation, subject to all individuals that follow the rule receiving a non-negative payoff while the unconditional defectors have $\mathrm{p}^*_i = 0$,
\begin{equation}
\begin{aligned}
\mathbf{p}^*  &= \underset{\mathbf{p} \ \in \ \left[0,1 \right]^N}{\text{arg max}}\left \{ \sum_{i\not\in\mathcal{D}} \mathrm{p}_i; \; \mathrm{y}_{i,\mathcal{G}}  \left( \mathbf{p}\right) \geq 0 \; \forall i \ \texttt{and} \ \mathrm{p}_i = 0 \; \forall i \in \mathcal{D} \right \}.
\end{aligned}
\label{eq:optimization-solution}
\end{equation}
In Appendix~B we provide further reasoning behind this representation. Moreover, there we also discuss that the solution to equation (\ref{eq:optimization-solution}) is unique, i.e. the corresponding dynamical system has a unique attractor. This implies that local cooperative behavior diffuses as a flux of energy over the network, eventually saturating at a point where cooperation is maximized under the constraint that no single individual is exploited.

\subsection{Phase transitions}

Next, we turn our attention to the phase transitions of the system. There are two such points which are of particular interest to us. The first transition corresponds to the situation when the behavioral rule  (\ref{eq:update}) is not able to support cooperation, i.e. below this point the steady state solution is described by $\mathbf{p}^* = \mathbf{0}$. 
The contrapositive of this condition represents a weak requirement for cooperation, in the sense that it represents a necessary condition for individuals with positive willingness for cooperation to exist.

The second transition point that we investigate quantifies a stronger condition for cooperation, corresponding to the situation where above this point all individuals (aside the unconditional defectors) are unconditional cooperators, i.e. the steady-state solution reads $\mathrm{p}^*_i = 1$ for all $i \not\in\mathcal{D}$. 

We study both transitions by using a differential equation representation of the update rule (see Appendix~C for more details regarding the conditions under which the representation is valid) 
\begin{align}
\dot{\mathrm{p}}_i =\dv{f_i \left( f_i^{-1} ( \mathrm{p}_i)\right)}{Y}  \left[ b_{i,\mathcal{G}} \left( \mathbf{p} \right) - c_{i,\mathcal{G}} \left( \mathrm{p}_i\right) \right].
\label{eq:differential}
\end{align}
In the following we provide a description of the conditions that are required for extinction and full unconditional cooperation to happen. The detailed proofs are provided in Appendix~C.

In particular, by analyzing the asymptotic stability of the system at the fixed point $\mathbf{p}^* = 0$ we show that the sufficient condition for extinction of cooperation can be approximated as
\begin{align}
\lambda_{\max} \left( \mathbf{J^{\setminus\mathcal{D}}_y}(\mathbf{0}) \right) < 0.
\label{eq:coop_extinciton_threshold}
\end{align}
where $\lambda_{\max}(\mathbf{J^{\setminus\mathcal{D}}_y}(\mathbf{p}))$ is the largest eigenvalue of~$\mathbf{J^{\setminus\mathcal{D}}_y}(\mathbf{p})$.

In fact, a similar result can be reached by studying the optimization problem~(\ref{eq:optimization-solution}). Altogether, this indicates that when every eigenvalue of the reduced Jacobian $ \mathbf{J^{\setminus\mathcal{D}}_y}(\mathbf{0})$ is negative, the system becomes dissipative and there is a continuous shrinkage in the displayed level of cooperation.

Similarly, by examining the stability of the linearized system at the point $\mathbf{1}_{\setminus\mathcal{D}}$ (the $N$ dimensional vector with entries $1$ for all $i \not\in \mathcal{D}$ and $0$ otherwise), we get that unconditional cooperation is asymptotically stable if 
\begin{align}
\min_{i \not\in \mathcal{D}} \left( v^{\setminus\mathcal{D}}_{i,\mathcal{G}} \right) > 1,
\label{eq:full_coop_threshold}
\end{align}
where $v^{\setminus\mathcal{D}}_{i,\mathcal{G}}$ is 
\begin{align}
v^{\setminus\mathcal{D}}_{i,\mathcal{G}} = \frac{b_{i,\mathcal{G}} (\mathbf{1}_{\setminus\mathcal{D}})}{ c_{i,\mathcal{G}}(1)}.
\end{align}

We note that the strict inequality in the approximation of the condition might be relaxed, as suggested by the sufficient and necessary conditions for unconditional cooperation. Indeed, on  the one hand, from the properties of the update rule we have that $v^{\setminus\mathcal{D}}_{i,\mathcal{G}} > 1$ for all $i \not\in \mathcal{D}$ implies $\mathrm{y}^*_{i,\mathcal{G}} > 0$, and hence $\mathrm{p}_i^* = 1$ for all $i \not\in \mathcal{D}$. On the other hand, in the necessary part the weak inequality $v^{\setminus\mathcal{D}}_{i,\mathcal{G}} \geq 1$ holds, which follows by directly substituting $\mathrm{p}_i^* = 1$ in (\ref{eq:payoff-general}) and repeating the same argument for all $i \not\in \mathcal{D}$. This further implies that individuals for which the inequality is not satisfied will never cooperate unconditionally. In this sense, the quantity $v^{\setminus\mathcal{D}}_{i,\mathcal{G}}$ arises as an index that quantifies the burden of each individual when cooperating, thus ultimately determining the degree of cooperation both at microscopic and macroscopic level. 

\section{Discussion}

\subsection{Public goods example}

As a constructive example for the applicative power of our framework, we consider an interaction mechanism which couples the iterated public goods game with a simple donation activity. 

In particular, we assume that in each time step nature randomly conditions whether the state of the global system is in provision of public goods or in donation. When in public goods state, each individual $i$ acts as a factor in $d_i$ processes of production represented by its nearest neighbors $\mathcal{N}^1_i$, with $d_i =|\mathcal{N}^1_i|$ being the degree of the individual. Its input in each round $t$ is proportional to the internal cooperative state $\mathrm{p}_i(t)$, whereas its productivity is inversely related with the degree $d_i$, thus leading to equal aggregate productivity of each individual.

For concreteness, we consider a linear production function,
\begin{align*}
\mathrm{q}_j(t) &= \alpha \sum_k \frac{A_{kj}}{d_k} \mathrm{p}_k ,
\end{align*}
where $\alpha$ is a parameter that describes the efficiency among the producers and $A_{kj} \in \{0,1\}$ is the $(k,j)$-th entry of the adjacency matrix $\mathbf{A}$ of the graph. In order to follow standard practice, we always fix $A_{ii} = 1$, thus each individual has one production process represented by its own vertex. To make the provision of public good related with generalized reciprocity we assume that after production takes place, $\mathrm{q}_j(t)$ is distributed to a randomly (on uniform) chosen individual from the nearest neighbors of $j$. 

On the other hand, when the system is in a donation state, each individual $i$ chooses a nearest neighbor at random and decides whether to donate him an amount $\alpha$ with probability proportional to its internal state $\mathrm{p}_i(t)$. 

\begin{figure}[t!]
\includegraphics[width=7.1cm]{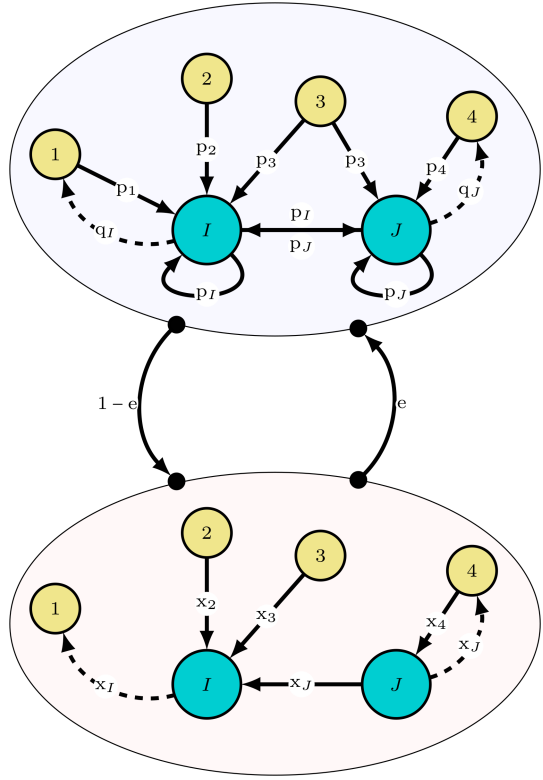} 
\caption{Scheme for the studied example. Top row shows one round of interactions for two production processes $I$ and $J$ in the network example that we examine. Solid directed edges indicate the contributions by individuals $i \in \{ 1,2,3,4 \}$, whereas the bended dashed edges are the randomly distributed outputs $q_j$, with $j \in  \{ I,J \}$. The bottom row describes the donation activity for the same individuals. }
\label{fig:gen-rep-example}
\end{figure}

The resulting public goods interaction is a randomized version of the model introduced in~\cite{santos2008social}, which has been extensively studied from the perspective of network reciprocity (a comprehensive review is provided  in~\cite{perc2013evolutionary}). In addition, when combined with the donation activity, it resembles the famous ``carrot'' mechanism in a generalized reciprocity system~\cite{milinski2012interaction}. One round of this interaction structure is depicted in Fig.~\ref{fig:gen-rep-example}.

\subsection{Properties}

In this structure, the random payoff of individual $i$ in round $t$ is defined as 
\begin{align*}
\mathrm{y}_i(t)&= \mathrm{e}(t) \mathrm{y}_{1i}(t) + (1 - \mathrm{e}(t)) \mathrm{y}_{2i}(t),
\end{align*}
where $\mathrm{e}(t)$ is a Bernoulli random variable with parameter $\varepsilon$ that describes the global state, and,
\begin{align}
\begin{split}
\mathrm{y}_{1i}(t) &= \sum_{j} \mathrm{m}_{1ji}(t) \mathrm{q}_j(t) - \mathrm{p}_i(t), \\
\mathrm{y}_{2i}(t) &= \alpha \sum_{j\neq i} \mathrm{m}_{2ji}(t) \mathrm{x}_j(t) - \mathrm{x}_i(t),
\end{split}
\label{eq:carrot-payoffs}
\end{align}
are the payoffs of the public goods game and the donation process. In equation (\ref{eq:carrot-payoffs}), $\mathrm{m}_{1ji}(t)$, $\mathrm{m}_{2ji}(t)$ and $\mathrm{x}_j(t)$ are, respectively, Bernoulli random variables with parameters $A_{ji} / d_j$, $A_{ji} / (d_j-1)$ and $\mathrm{p}_j(t)$.  

In this example, the benefit and cost functions represent affine maps with respect to the random variables. Therefore, we can approximate the analysis of the steady state with the following deterministic payoff
\begin{widetext}
\begin{eqnarray}
\mathrm{y}_i(t) = \varepsilon \alpha \sum_j \frac{A_{ji}}{d_j} \sum_k \frac{A_{kj}}{d_k} \mathrm{p}_k(t) + (1- \varepsilon) \alpha \sum_{k\neq i} \frac{A_{ki}}{d_k-1} \mathrm{p}_k(t) - \mathrm{p}_i(t).
\label{eq:carrot-deterministic}
\end{eqnarray}
\end{widetext}

From (\ref{eq:carrot-deterministic}) one can easily deduce that for each particular individual $i$, conditions (\ref{eq:dilemma_condition_1}) and  (\ref{eq:dilemma_condition_2}) are satisfied whenever $\alpha \in \left( 1, \frac{d_i}{z_{i}} \frac{1}{\varepsilon}\right)$, where  $z_{i} = \sum_j A_{ji}/d_j$ is a centrality measure which quantifies the expected number of times that individual $i$ is selected to be at the receiving end of a production process or a donation. The effect of the distribution of $z_{i}$ (also referred to as neighborhood importance index) on the network propensity for cooperation has been discussed in \cite{Utkovski-2017}.  

A refined version of this index, which is proportional to the maximal average benefit of individual $i$, can be defined as 
\begin{align}
s^{\setminus\mathcal{D}}_{i} = \varepsilon \sum_j \frac{A_{ji}}{d_j} \sum_{k\not\in\mathcal{D}} \frac{A_{kj}}{d_k} + (1 - \varepsilon) \sum_{k\not\in\mathcal{D}\cup i} \frac{A_{ki}}{d_k-1}.
\label{eq:pgodds-index}
\end{align}
 Its relation with the efficiency parameter $\alpha$ directly determines the level of cooperation of the system because the cooperation index $v^{\setminus\mathcal{D}}_{i,\mathcal{G}}$ can be written as 
\begin{align*}
v^{\setminus\mathcal{D}}_{i,\mathcal{G}} &= \alpha s^{\setminus\mathcal{D}}_{i}.
\end{align*}
This implies that full cooperation will exist whenever
\begin{align}
\alpha > \frac{1}{s^{\setminus\mathcal{D}}_{\min}},
\label{eq:full-coop-pgoods}
\end{align}
where $s^{\setminus\mathcal{D}}_{\min}$ is the minimum among the $s^{\setminus\mathcal{D}}$ indices.

Finally, the matrix $\mathbf{J^{\setminus\mathcal{D}}_y}(\mathbf{0})$ whose largest eigenvalue determines whether cooperation will die out or not is 
\begin{align*}
\mathbf{J^{\setminus\mathcal{D}}_y} (\mathbf{0}) &= \alpha \left[\varepsilon \mathbf{M_1^{\setminus\mathcal{D}}} + (1 - \varepsilon)\mathbf{M_2^{\setminus\mathcal{D}}} \right] - \mathbf{I},
\end{align*}
where the $(i,j)$-th entries of $\mathbf{M_1^{\setminus\mathcal{D}}}$ and $\mathbf{M_2^{\setminus\mathcal{D}}}$ are $M^{\setminus\mathcal{D}}_{1ij} = \sum_k \frac{A_{ki}}{d_k} \frac{A_{jk}}{d_j}$ and $M^{\setminus\mathcal{D}}_{2ij} = \frac{A_{ji}}{d_j-1}$ ( $M^{\setminus\mathcal{D}}_{2ii} = 0$), and $\mathbf{I}$ is the identity matrix.

This implies that the condition for extinction of cooperation reads
\begin{align}
\alpha < \frac{1}{\lambda_{\max}\left(\varepsilon\mathbf{M_1^{\setminus\mathcal{D}}} + (1 - \varepsilon)\mathbf{M_2^{\setminus\mathcal{D}}}\right)}.
\label{eq:extinction-coop-pgoods}
\end{align}

In the special case when every individual follows the update (\ref{eq:update}), $\mathbf{M_1^{\setminus\mathcal{D}}}$ and $\mathbf{M_2^{\setminus\mathcal{D}}}$ represent left stochastic matrices, in which case  $\lambda_{\max}\left(\varepsilon\mathbf{M_1^{\setminus\mathcal{D}}} + (1 - \varepsilon)\mathbf{M_2^{\setminus\mathcal{D}}}\right)  = 1$. Since $\alpha > 1$ is prerequisite for a social dilemma, this condition will never hold in a population consisting exclusively of individuals that follow the state-based rule (i.e. without defectors).

\subsection{Implications}

 We test these analytical findings on three different types of random graph models, the Random Regular (RR) graph, the Erdos-Renyi (ER) random graph and the Barabasi-Albert (BA) scale-free network. 
 
\begin{figure*}[t!]
\begin{adjustwidth}{-0.2in}{0in}
\includegraphics[width=17cm]{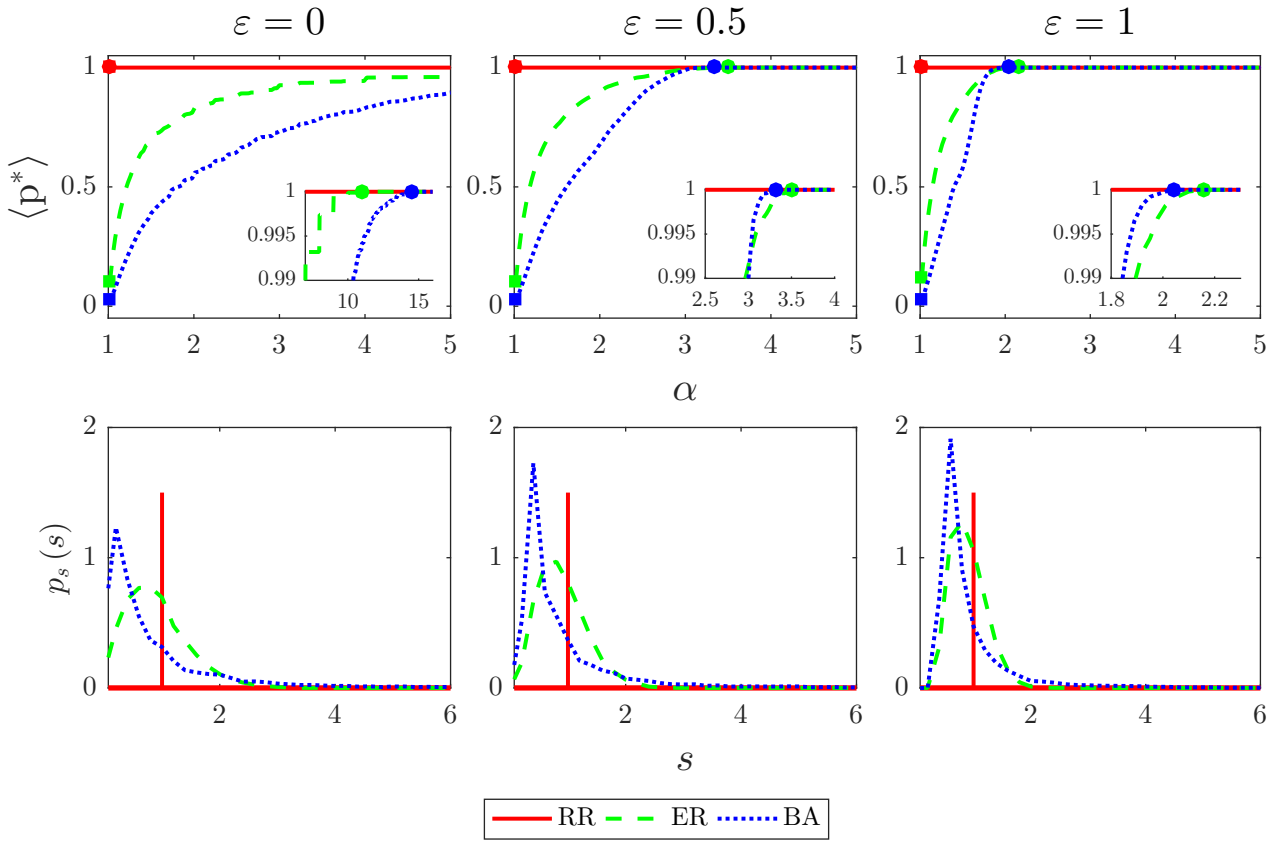}
\caption{Results for the public goods example. 
Top row shows the fraction of unconditional cooperators $\langle \mathrm{p}^* \rangle$ as a function of $\alpha$, while the bottom row gives the estimated probability density function of the index $s$. The columns correspond to different values of the parameter $\epsilon$. The results are averaged over 100 realizations with each graph having $N = 100$ and average degree $5$.
\label{fig:initial_results}}
\end{adjustwidth}
\end{figure*}

Throughout the analysis, as a measure for the global level of cooperation we consider the fraction of unconditional cooperators $\langle \mathrm{p}^* \rangle$ out of the individuals that follow the state-based update rule. In the top row of Fig.~\ref{fig:initial_results} we study the evolution of this variable as a function of the efficiency parameter $\alpha$ when $\mathcal{D} = \emptyset$ and by considering three cases of $\varepsilon$. Namely, we examine the situations when there is only a donation process~($\varepsilon = 0$), when there is equal probability for happening of both processes~($\varepsilon = 0.5$) and when only public goods are present in the system~($\varepsilon = 1$).

The condition for extinction of cooperation is plotted in the figure as a square point. 
As suggested by the analytical findings, cooperation in each graph and interaction structure exists as long as $\alpha$ satisfies the condition  for a social dilemma. 

However, the dependency of $\langle \mathrm{p}^* \rangle$ on $\alpha$ is generally disparate across the three interaction structures we explore, with the only similarity arising in the RR graph. In that case, it can be seen that full unconditional cooperation happens always independently of the interaction structure. For the ER and BA graphs, we discover that in each interaction structure the former graph acts as a better promoter of cooperation for low values of $\alpha$. While this persists when donation is the sole mechanism, whenever provisioning of public goods enters the system, a critical point appears beyond which the BA graph performs better in terms of the number of unconditional cooperators (see the inset plots).

To interpret these observations, in the bottom row of Fig.~\ref{fig:initial_results} we display the probability density function of the index $s^{\setminus\mathcal{D}}$, when there are no defectors (here simply denoted as $s$). The figure shows that the distribution of this variable in the RR graph resembles a Dirac delta function with the mass centered on $1$. This implies that the condition for asymptotic stability of unconditional cooperation is satisfied for every plausible $\alpha$, which in turn implies the full unconditional cooperation observed at $\alpha = 1$. For the ER and BA random graphs we notice that the distribution of $s$ is positively skewed with the BA graph exhibiting by far larger skewness. A direct consequence of this is the remark for the evolution of $\langle \mathrm{p}^* \rangle$ in the two graphs. More precisely, the right tail of the distribution determines the degree of cooperation provided when the level of efficiency is low. In this regard the lower slope in the ER graph indicates that for low $\alpha$, there will be more individuals for which the necessary condition for unconditional cooperation is satisfied.

Similarly, the left tail explains why the BA graph converges to full unconditional cooperation faster than the ER graph in the case of public goods. In particular, as $\varepsilon$ increases the kurtosis in the distribution of $s$ also increases for both ER and BA graphs, implying that the individuals tend to become more similar in regard to this index. The increment is larger for the BA graph, and therefore the lower thresholds for unconditional cooperation in the case of public goods.

\begin{figure*}[t!]
\begin{adjustwidth}{-0.1in}{0in}
\includegraphics[width=17cm]{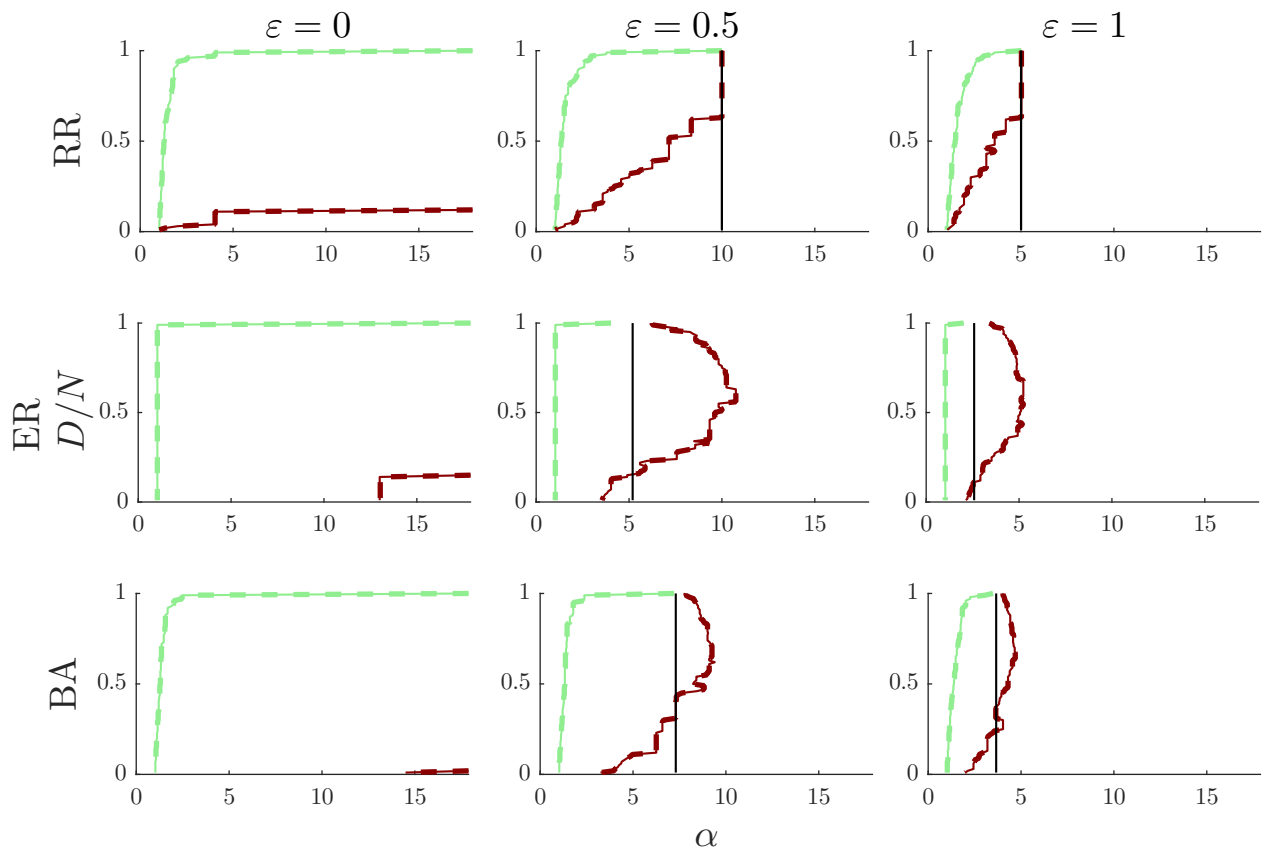}
\caption{Contour plots for the transition points. The green (light gray) curves are contour plots for the transition points to extinction of cooperation while the red (dark gray) curves are the transition points to full unconditional cooperation, with the dashed lines indicating the estimated thresholds from equations (\ref{eq:coop_extinciton_threshold}) and (\ref{eq:full_coop_threshold}). With the black vertical line we denote the minimum of $d/z$. The rows represent different types of random graphs (RR, ER and BA) while the columns correspond to different values of $\epsilon$. The results are averaged across 100 graph realizations consisting of 100 individuals with each graph having an average degree of $5$.}
\label{fig:results_defectors}
\end{adjustwidth}
\end{figure*}

In order to quantify the effect of defectors, in Fig.~\ref{fig:results_defectors} we display contour plots for the fraction of unconditional cooperators as a function of both the efficiency parameter and the fraction of individuals that are unconditional defectors for each of the studied random graphs. In the figure, above the green (light gray) lines is the region where cooperation ceases to exist, whereas below the red (dark gray) lines is the region where all individuals outside the defector set cooperate unconditionally. In the region in-between, unconditional cooperators coexist with partial cooperators (those with $0 < \mathrm{p}^*_i < 1$) and unconditional defectors. As defectors we always set the $D$ individuals with the highest values of $d/z$. This is the group of individuals for which the social dilemma requires a larger $\alpha$ as a means to disappear, if there is a positive probability for reaching a public goods state.

We find that, among the random graph models, the ER graph model requires the lowest efficiency (benefit to cost ratio) for the cooperation to persist (i.e. not to be extinct), followed by the RR graph. The BA graph presents itself as the topology where extinction of cooperation is more probable (compared with ER and RR graphs), for the same average graph degree. We remark that this observation is independent on the choice for $\varepsilon$. In contrast, it can be noticed that there is quite a colorful discrepancy with respect to the threshold above which full (i.e. unconditional) cooperation begins. In this aspect, when donation is the only natural process, the RR graph is the most supportive to cooperation, followed by the ER and the BA graph. However, when public goods is included in the possible natural states, it seems that when around half of the population is constituted by defectors the BA graph performs better in promoting unconditional cooperation, while the ER graph is the best promoter of unconditional cooperation when the majority of the population are defectors. 

Similarly to the scenario without defectors, in the case with defectors the findings can also be attributed to the distribution of the index $s$. Concretely, the inclusion of defectors can be described as an internal force which ultimately decreases the value of $s$. From this point of view, the extinction of cooperation is related to the right tail of the distribution since then one faces instances where the necessary condition for existence of cooperation is not fulfilled. As discussed previously, the ER graph has the largest tail for the distribution of the index $s$ among the studied random graphs, and therefore is most robust to extinction of cooperation when defectors are included. On the other hand, it seems considerably more delicate to precisely quantify the differences between the random graph models with respect to the existence of full unconditional cooperation as function of the distribution of the index $s$, as the results depend on the specific choice of the individuals which are selected as defectors. Nonetheless, one can notice that, as the number of defectors increases the distribution of $s$ is driven towards lower values and thus the threshold for full cooperation increases. Evidently this has lowest effect on the RR graph since the distribution of $s$ in that case concentrates around one value, i.e. there are no tails. Hence the observed lower thresholds for full unconditional cooperation in the sole donation case.  We note that, in general, this explanation does not hold when provision of public goods is a possible state in the global system since then, as shown previously, for each individual a critical point exists after which cooperation is no longer a social dilemma. For instance, the value of this critical point in the RR graph is independent of the selected individual (its minimum for each graph and structure in the figure is denoted as a black vertical line). From the figure, it appears that the ER graph requires the lowest efficiency in order to have a nonempty set of individuals for which the social dilemma disappears, followed by the BA graph. A direct corollary of this is the observed lower threshold for full unconditional cooperation in these two graphs, compared to the RR graph, when a large fraction of the population behaves as defectors.

\section{Conclusion} 

In this paper we developed a unifying framework for studying the role of a state-based behavioral mechanism centered on generalized reciprocity, on the cooperation dynamics in complex networks. Such mechanisms have been recently discovered in natural systems which involve low-level interactions among network agents with limited processing/cognitive abilities. Interestingly, they have also been attested at the more intelligent level of human interactions~\cite{Bartlett-2006,Stanca-2009}. 

While there is significant empirical evidence for the presence of these mechanisms in a plethora of real-life systems, the role of the network structure on the cooperation dynamics under general interaction models has only recently start to shape the research in various fields. In this context, we believe that the here introduced framework provides a systematic way to study the various aspects of cooperation in complex networks. In particular, the generality of the addressed framework allows for incorporation of a wide range of social dilemmas and interaction structures in the model. 

Besides its theoretical value, we believe that the framework may also be used to quantify the cooperation dynamics in real-life systems governed by similar behavioral mechanisms. We refer to a recent experimental study which suggests that the pay-it-forward principle of generalized reciprocity is a better promoter of long-term cooperation among humans than indirect reciprocity simply because it is cognitively less demanding~\cite{baker2014paying}. In this context, our (and similar) models may provide the theoretical background for the observed long-term behavior. Indeed, the above observation may be addressed from the dynamical systems perspective, under which the application of a behavioral mechanism based on generalized reciprocity yields a unique attractor where the overall level of cooperation is maximized, while at the same time the involved individuals are prevented from exploitation.

As a final remark, we point out that the presented framework may apply beyond the studied structures and can easily be coupled with other mutual processes found in social and natural systems~\cite{nicosia2017collective,antonioni2017coevolution}. As such it acts as a building block for investigating the role of state-based generalized reciprocity in the dynamics of intertwining phenomena of different nature. This represents an interesting direction for future work. 

\section*{Acknowledgement}

This research was supported in part by DFG through grant ``Random search processes, L\'evy flights, and random walks on complex networks''.

\subsection*{Appendix A: Social dilemma examples}
\setcounter{equation}{0}
\setcounter{figure}{0}
\setcounter{table}{0}
\setcounter{theorem}{0}
\makeatletter
\renewcommand{\theequation}{A\arabic{equation}}
\renewcommand{\thetable}{A\arabic{table}}
\renewcommand{\thefigure}{A\arabic{figure}}
\renewcommand{\thetheorem}{A\arabic{theorem}}
\renewcommand{\theproposition}{A\arabic{proposition}}

\textbf{The Prisoner's Dilemma:} This is the most famous example for a social dilemma where the interactions are pairwise. In the simplest case, in a prisoner's dilemma a cooperator pays a cost $\gamma$ for the other individual to receive a benefit $\beta$ \cite{Nowak-2006five}. In the original case, $\beta > \gamma$ is a prerequisite in order for a social dilemma to exist. When extended to a network structure, each individual meets with its nearest neighbors $\mathcal{N}^{1}_i$ and experiences the same interactions. Therefore, the payoff of individual $i$ in a prisoner's dilemma played on a network where the strategies are formed by a state-based update can be written as
\begin{align}
\mathrm{y}_i(t) &= \beta \sum_j A_{ij} \mathrm{p}_j(t) - \gamma d_i \mathrm{p}_i(t).
\label{eq:pris-dilemma}
\end{align}
To exist a dilemma, equation (\ref{eq:dilemma_condition_1}) needs to hold for all $\mathbf{\hat{p}}$ and equation (\ref{eq:dilemma_condition_1}) has to be a strict inequality for all $\mathbf{\hat{p}}$. It is obvious that this will happen if $\beta > \gamma$, as in the original interaction structure.

The prisoner's dilemma can be easily extended to account for random interactions, and thus to resemble more closely interactions where generalized reciprocity can happen. For example, this is done in references~\cite{Barta-2011,Utkovski-2017}.

\textbf{Snowdrift Game:} The snowdrift game is the simplest form of a nonlinear interaction structure where cooperation by an individual is an advantageous strategy whenever the state of the neighboring individuals is defective. 

The payoff in a pairwise snowdrift game defined on a network, can be written in a similar way as in the prisoner's dilemma. In particular, the payoff of individual $i$ in round $t$ is 
\begin{widetext}
\begin{align}
\mathrm{y}_i(t) &= \sum_j A_{ij} \left[ \beta \mathrm{p}_j(t) + \beta \mathrm{p}_i(t)  + \mathrm{p}_i(t)(\frac{\gamma}{2} - \beta)\mathrm{p}_j(t) \right] - \gamma d_i \mathrm{p}_i(t).
\label{eq:snowdrift-pairwise}
\end{align}
\end{widetext}
In this interaction structure, condition (\ref{eq:dilemma_condition_1}) implies that cooperation by $i$ will be a favorable threat as long as the average level of cooperation by the neighbors satisfies the following inequality
\begin{align*}
\sum_j \frac{A_{ij}}{d_i} \mathrm{p}_j < \frac{\beta - \gamma}{\beta - \gamma/2 }.
\end{align*}

On the other hand, condition (\ref{eq:dilemma_condition_2}) will be satisfied whenever $\beta > \gamma/2$.

Interestingly, one may find out that, by applying the rules for extinction and full cooperation that we described, a state-based generalized reciprocity update rule always leads to full cooperation by the individuals that follow it as long as $\beta > \gamma$. This result is independent of the structure of unconditional defectors.

\textbf{Common-pool Resource problem:} Common-pool resource problem is a situation that best depicts the ``tragedy of the commons''~\cite{Hardin-2009}. Formally, a tragedy of commons is a situation where there is shared renewable resource that will continue to produce benefits if the population does not over harvest it but in which any single individual profits from harvesting as much as possible~\cite{Brechner-1977}.

As a typical example for this situation we consider a model where a mixed population (represented by a fully connected network) at each round $t$ are harvesting a resource $R(t)$ with an amount that is inversely proportional to their internal state. The payoff of each individual $i$ is given as
\begin{align}
\mathrm{y}_i(t) &= (1 - \delta \mathrm{p}_i(t))(R(t) - 1) - \gamma \mathrm{p}_i(t),
\label{eq:cpr}
\end{align}
where $\delta$ is a parameter which regulates the optimal consumption by the group (optimally each individual would consume $(1-\delta)$ fraction of the resource, and $\gamma$ is the opportunity cost. 

Different dynamics can be used for the evolution of $R(t+1)$. For simplicity we assume that in each round the resource grows by one unit and depletes by the average amount of the consumption by all individuals (for other representations see~\cite{sugiarto2017social}). Mathematically, this is written as
\begin{align}
R(t+1) &= R(t) + 1 - \frac{R}{N} \sum_j \left(1 - \delta \mathrm{p}_j(t) \right).
\label{eq:cpr-resource}
\end{align}
To estimate $\mathbf{p}^*$, one first needs to estimate the equilibrium resource $R^*$ from equation (\ref{eq:cpr-resource}) and substitute the result in equation (\ref{eq:cpr}).

Then, the conditions (\ref{eq:dilemma_condition_1}) and (\ref{eq:dilemma_condition_2}) for social dilemma are given as $0 < \gamma < \delta < 1$ and $\delta < \gamma N$.

\subsection*{Appendix B: Steady state analysis}
\setcounter{equation}{0}
\setcounter{figure}{0}
\setcounter{table}{0}
\setcounter{theorem}{0}
\makeatletter
\renewcommand{\theequation}{B\arabic{equation}}
\renewcommand{\thetable}{B\arabic{table}}
\renewcommand{\thefigure}{B\arabic{figure}}
\renewcommand{\thetheorem}{B\arabic{theorem}}
\renewcommand{\theproposition}{B\arabic{proposition}}

In the main text, we pointed out two assumptions on which our analysis is built. First, in every case we consider a fraction of individuals that behave as unconditional defectors, i.e. $\mathrm{p}_i(t) = 0$ for all $t$. This indicates that we can reduce the examination of the model to the properties of the payoff functions of each $i \not\in  \mathcal{D}$ defined on the set ~$\mathcal{P}^N_\mathcal{D}~= ~\left\{ \mathbf{p}_{\mathcal{D}} \in \left[ 0,1 \right]^N : \forall i \in \mathcal{D}, \ \mathrm{proj}_i (\mathbf{p}_{\mathcal{D}}) = 0  \right\}$. Moreover, we assumed that the interaction structure is non-degenerate, i.e. $\mathbf{J_y}^{\setminus\mathcal{D}}(\mathbf{0})$, where $\mathrm{J_y}^{\setminus\mathcal{D}}_{ij}(\mathbf{0}) =  \pdv{\mathrm{y}_{i,\mathcal{G}}(\mathbf{0})}{\mathrm{p}_j}$ for all $i,j \not\in\mathcal{D}$, is a nonsingular matrix. As we will see, the second assumption ensures that the system will never be stuck in a saddle point. Against this background, we can derive the described proposition for the steady state solution.

\begin{proposition}
-- \textbf{Steady State Solution:} If a fraction of the individuals base their state update $\mathrm{p}_i(t)$ on the rule 
\begin{align*}
\mathrm{p}_i(t+1)=f_i \left(\mathrm{Y}_{i,\mathcal{G}}(t)\right),
\end{align*}
where $\mathrm{Y}_{i,\mathcal{G}}(t)=\mathrm{Y}_{i,\mathcal{G}}(t-1)+\mathrm{y}_{i,\mathcal{G}}(t)$, with $\mathrm{Y}_{i,\mathcal{G}}(0)$ being the initial condition and $\mathrm{y}_{i,\mathcal{G}}(0)=0$, and $f_i :\mathbb{R} \to \left(0, 1 \right)$, is a strictly monotonic homeomorphism that satisfies $\lim_{\mathrm{Y}\to -\infty} f_i(\mathrm{Y}) = 0$ and $\lim_{\mathrm{Y}\to \infty} f_i(\mathrm{Y}) = 1$, then the stable steady state solution $\mathbf{p}^*$ is unique and can be found as the solution to the constrained optimization problem of maximizing the overall level of cooperation such that every individual has a nonnegative payoff,~i.e.,
\begin{equation}
\begin{aligned}
\mathbf{p}^*  &= \underset{\mathbf{p} \ \in \ \mathcal{P}^N_\mathcal{D}}{\text{arg max}}\left \{ \sum_{i\not\in\mathcal{D}} \mathrm{p}_i; \; \mathrm{y}_{i,\mathcal{G}}  \left( \mathbf{p}\right) \geq 0 \; \forall i\right \}.
\end{aligned}
\label{eq:optimization-solution-supplementary}
\end{equation}
\end{proposition}

\textit{Sketch of proof:} For all $i \not\in \mathcal{D}$, fix $\mathrm{Y}_{i,\mathcal{G}}(0)$ to some arbitrary real number. By construction, $\mathrm{p}_i(1)~\in~\left(0,1\right)$. Moreover, the  reduced individual payoff vector $\mathbf{y}_{\setminus\mathcal{D}}(t)$ that accounts for the payoffs of all those $i$ can be approximated as
\begin{align*}
\mathbf{y}_{\setminus\mathcal{D}}(1) &\approx \mathbf{J^{\setminus\mathcal{D}}_y}(\mathbf{0}) \cdot \mathbf{p}_{\setminus\mathcal{D}}(1),
\end{align*}
where $\mathbf{p}_{\setminus\mathcal{D}}(1)$ is the corresponding state vector in $t = 1$.

Since $\mathbf{J^{\setminus\mathcal{D}}_y}(\mathbf{0})$ is nonsingular, there exists some $i$ in the set of individuals that follow the rule such that $\mathrm{y}_i(1) \neq 0$. Without loss of generality, let us assume that for all those $i$ $\mathrm{y}_{i,\mathcal{G}}(1) > 0$ (similar arguments hold when there exist individuals with $\mathrm{y}_{i,\mathcal{G}}(1) < 0$). The strict monotonicity implies that these individuals will increase their cooperative state in the following period.
By iterating the same procedure for a finite amount of times we can easily deduce that similar dynamics will appear in every period $t$ (the set of individuals with positive payoffs may change).
In the limit as $t \to \infty$, the individuals with positive payoffs will be able to increase their states until $\mathrm{p}^*_i = 1$ or some $\mathrm{p}^*_i$ that satisfies $\mathrm{y}_{i,\mathcal{G}}(\mathbf{p}^*) = 0$. The second possibility is again a result of the strict monotonicity of $f$ (a negative payoff implies that the individual will decrease its cooperative state until the condition is reached). Therefore, in the steady state each individual conforming to (\ref{eq:update}) will cooperate with the maximal willingness, based on its environment, so as it is not exploited, indicating that the steady state internal cooperative state vector $\mathbf{p}^*$ can be found as a solution to~(\ref{eq:optimization-solution-supplementary}).

Notice that a steady state solution can still be found as a solution to (\ref{eq:optimization-solution-supplementary}) even if $\mathbf{J_y}(\mathbf{0})$. However, then due to the singularity of $\mathbf{J_y}(\mathbf{0})$ multiple steady states may appear that behave as saddle points.

Now let us turn to the uniqueness part. It is known that a constrained optimization problem will have a unique maximum if the objective function (in this case $\sum_i \mathrm{p}_i$ ) is strictly quasiconcave and all constraints are quasiconcave ($\mathrm{y}_i(\mathbf{p})$). Formally, a continuously differentiable function $\mathrm{f}$ is quasiconcave in a convex set $\mathcal{X}$, if for any two points $\mathrm{x}_1$ and $\mathrm{x}_2$ in the set such that $\mathrm{f}(\mathbf{x}_1) \geq \mathrm{f}(\mathbf{x}_2)$ we have that 
\begin{align}
\mathbf{J}_\mathrm{f} \left( \mathbf{x}_2 \right) \left( \mathbf{x}_1 - \mathbf{x}_2 \right) \geq 0,
\label{eq:concavity}
\end{align}
where $\mathbf{J}_\mathrm{f} \left( \cdot \right)$ is the Jacobian of the function $\mathrm{f}$, with the strict inequality whenever implying that the function is strictly quasiconcave. Further information about the properties of (strictly) quasiconcave functions and their role in constrained optimization can be found in \cite{de2000mathematical}.

For simpler notation, let $\mathrm{p}_{\Sigma}(\mathbf{p}) \doteq \sum_{i\not\in\mathcal{D}} \mathrm{p}_i$. Since $\mathrm{p}_{\Sigma}(\mathbf{p})$ is linear with respect to each projection, it is easy to show that for every $\mathbf{p}_1$ and $\mathbf{p}_2$ such that 
$\mathrm{p}_{\Sigma}(\mathbf{p}_1) > \mathrm{p}_{\Sigma}(\mathbf{p}_2)$, the strict inequality in (\ref{eq:concavity}) will hold. In addition, by Taylor expanding $\mathrm{y}_i (\mathbf{p})$ for all $i$, at a point $\mathbf{p}_2$ it follows that the constraints are quasiconcave for all $\mathbf{p}_1$ such that $\mathrm{y}_i (\mathbf{p}_1) \geq \mathrm{y}_i (\mathbf{p}_2)$.  \hfill $\blacksquare$

\subsection*{Appendix C: Phase transition analysis}
\setcounter{equation}{0}
\setcounter{figure}{0}
\setcounter{table}{0}
\setcounter{theorem}{0}
\makeatletter
\renewcommand{\theequation}{C\arabic{equation}}
\renewcommand{\thetable}{C\arabic{table}}
\renewcommand{\thefigure}{C\arabic{figure}}
\renewcommand{\thetheorem}{C\arabic{theorem}}
\renewcommand{\theproposition}{C\arabic{proposition}}

The analysis for the phase transitions of the system relies heavily on the differential equation form (\ref{eq:differential}). Therefore, before we analyze them we rewrite the discrete system in the continuous form presented in the main text. Concretely, the update rule can be described with the following recurrence relation
\begin{widetext}
\begin{align}
\mathrm{p}_i(t+1) &= f_i\big(f_i^{-1} \left( \mathrm{p}_i(t)\right)+b_{i,\mathcal{G}} \left( \mathbf{p}(t) \right) - c_{i,\mathcal{G}} \left( \mathrm{p}_i(t) \right)\big).
\label{eq:iterative_formulation}
\end{align}
\end{widetext}
Note that from this representation the steady state equation (\ref{eq:iterative_steady_state}) is derived.

By expanding equation (\ref{eq:iterative_formulation}) point-wise through Taylor series at the point $ f_i^{-1} \left( \mathrm{p}_i(t)\right)$ we can approximate $\mathrm{p}_i(t+1)$ as
\begin{widetext}
\begin{align*}
\mathrm{p}_i(t+1) \approx \mathrm{p}_i(t) + \dv{f_i \left( f_i^{-1} ( \mathrm{p}_i(t))\right) }{Y}  \left[ b_{i,\mathcal{G}} \left( \mathbf{p}(t) \right) - c_{i,\mathcal{G}} \left( \mathrm{p}_i(t) \right) \right],
\end{align*}
\end{widetext}
which can be written in the differential form given in equation (\ref{eq:differential}), i.e.,
\begin{align*}
\dot{\mathrm{p}_i} = \dv{f_i \left( f_i^{-1} ( \mathrm{p}_i)\right)}{Y} \left[ b_{i,\mathcal{G}} \left( \mathbf{p} \right) - c_{i,\mathcal{G}} \left( \mathrm{p}_i\right) \right],
\end{align*}
for all $i \not\in \mathcal{D}$.

Moreover, as can be seen from equation (\ref{eq:differential}) the study of the system is dependent on the properties of the (higher order) derivatives of $f_i$. Since we do not have a specific form for the update rule, they may be difficult to attain. Nevertheless, we can use a simple trick based on the smoothness of the update rule to discover the properties. In particular, notice that the famous Logistic function, defined as 
\begin{align*}
g(\mathrm{Y})=\left[1+e^{-\kappa(\mathrm{Y}-\omega)}\right]^{-1},
\end{align*}
where the parameters $\kappa$ and $\omega$ define the steepness, respectively the midpoint of the function, is a specific function that satisfies our definition for being a generalized reciprocity update rule. More importantly, this function has well-defined derivatives. For example the first derivative is $\dv{g\left(\mathrm{\hat{Y}} \right)}{\mathrm{Y}} = g(\mathrm{\hat{Y}}) \left( 1 - g(\mathrm{\hat{Y}}) \right)$ and the second is $\dv[2]{g\left(\mathrm{\hat{Y}} \right)}{\mathrm{Y}} = g(\mathrm{\hat{Y}}) \left( 1 - g(\mathrm{\hat{Y}}) \right) \left( 1 - 2 g(\mathrm{\hat{Y}}) \right)$. 

\begin{figure}[t!]
\includegraphics[width=8.6cm]{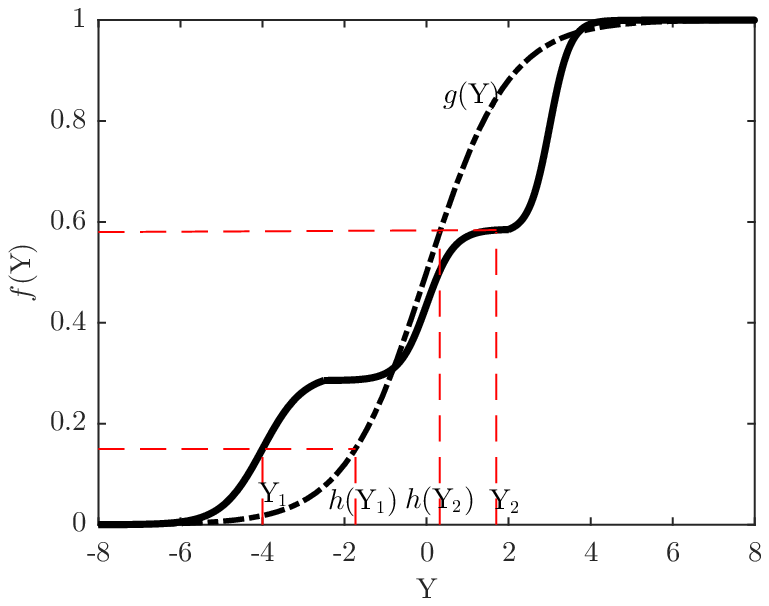}
\caption{An example for monotonic transformation of $f$. Here, the accumulated payoffs $\mathbf{Y}_j$ for $j \in \{1,2 \}$ are mapped to $h(\mathbf{Y}_j)$ so as $g \left(h(\mathbf{Y}_j)\right) = f(\mathbf{Y}_j)$.} 
\label{fig:monotonic_transformation}
\end{figure}

Hence, we can define a continuously differentiable and monotonically increasing function $h_i:\mathbb{R} \to \mathbb{R}$ that acts as a monotonic transformation of the accumulated payoff and which satisfies
\begin{align}
f_i(\mathrm{Y}) &= g(h_i(\mathrm{Y})).
\label{eq:update_chain_rule}
\end{align}
Besides this, the function $h_i$ also satisfies $\lim_{p_i \to 1} h_i ( f_i^{-1}(p_i)) = f_i^{-1}(p_i)$ and ~$\lim_{p_i \to 0} h_i ( f_i^{-1}(p_i)) = f_i^{-1}(p_i)$. An illustrative example for this transformation is given in Fig.~\ref{fig:monotonic_transformation}.
Now, studying the derivatives of $f_i$ is a simpler task since we can implement the chain rule and use the properties of the derivatives of the Logistic function. 

\vspace{1cm}

\begin{proposition}
-- \textbf{Extinction of cooperation:} If 
\begin{align}
\lambda_{\max} ( \mathbf{J^{\setminus\mathcal{D}}_y}(\mathbf{0})) & < 0,
\end{align}

where $\lambda_{\max}(\mathbf{J^{\setminus\mathcal{D}}_y}(\mathbf{0}))$ is the largest eigenvalue of the matrix $\mathbf{J^{\setminus\mathcal{D}}_y}(\mathbf{0})$ with entries defined as $\mathrm{J_y}_{ij}(\mathbf{0}) =  \pdv{\mathrm{y}_i(\mathbf{0})}{\mathrm{p}_j}$ for all $i,j \not\in\mathcal{D}$, then $\mathbf{p}^* = 0$ is an asymptotically stable steady state.
\end{proposition}

\textit{Proof:} To prove this proposition we implement Lyapunov's direct method for stability. First, we define the function $L(\mathbf{p}) = \sum_{i\not\in\mathcal{D}} \mathrm{p}_i$. It is clear that this function is a candidate for being a Lyapunov function since $L(\mathbf{0}) = 0$ and $L(\mathbf{\hat{p}}) > 0$ for all $\mathbf{\hat{p}} \neq \mathbf{0}$. 

The time derivative of $L(\mathbf{p})$ is
\begin{widetext}
\begin{align}
\dot{L}(\mathbf{p}) &= \sum_{i \not\in \mathcal{D}} \mathrm{p}_i \left(1 - \mathrm{p}_i  \right) \dv{h_i\left( f_i^{-1} ( \mathrm{p}_i)\right)}{Y} \left[ b_{i,\mathcal{G}} \left( \mathbf{p} \right) - c_{i,\mathcal{G}} \left( \mathrm{p}_i\right) \right], 
\label{eq:lyapunov-ineq}
\end{align}
\end{widetext}
which in a neighborhood around $\mathbf{p} = 0$ can be approximated as
\begin{align}
\dot{L}(\mathbf{p}) &\approx \sum_{i \not\in \mathcal{D}} \mathrm{p}_i \left[ \sum_{j \not\in \mathcal{D}} \pdv{b_{i,\mathcal{G}} (\mathbf{0})}{\mathrm{p}_j} \mathrm{p}_j - \dv{c_{i,\mathcal{G}} (0)}{\mathrm{p}_i} \mathrm{p}_i\right], \\
&= \mathbf{p_{\setminus\mathcal{D}}}^T\mathbf{J^{\setminus\mathcal{D}}_y}(\mathbf{0})\mathbf{p_{\setminus\mathcal{D}}},
\label{eq:taylor-lyapunov-ineq}
\end{align}
where $T$ is the transpose operator. For the system to be asymptotically stable at $\mathbf{p}^* = \mathbf{0}$, expression (\ref{eq:taylor-lyapunov-ineq}) has to be negative for all $\mathbf{p} \neq \mathbf{0}$. Clearly, (\ref{eq:taylor-lyapunov-ineq}) is a quadratic form, therefore this will happen only if $\mathbf{J^{\setminus\mathcal{D}}_y}(\mathbf{0})$ is negative definite, i.e. if $\lambda_{\max} (\mathbf{J^{\setminus\mathcal{D}}_y}(\mathbf{0})) < 0$. \hfill $\blacksquare$

\begin{proposition}
-- \textbf{Unconditional cooperation:} Unconditional cooperation, defined as $\mathbf{p}^* = \mathbf{1}_{\setminus\mathcal{D}}$, where $\mathbf{1}_{\setminus\mathcal{D}}$ is an $N \times 1$ dimensional vector with entries $1$ for all $i \not\in  \mathcal{D}$ and $0$ otherwise, is an asymptotically stable steady state if
\begin{align*}
\min_{i \not\in \mathcal{D}} \left( v^{\setminus\mathcal{D}}_{i,\mathcal{G}}  \right) > 1,
\end{align*}
\end{proposition}
where $v^{\setminus\mathcal{D}}_{i,\mathcal{G}} = \frac{b_{i,\mathcal{G}} (\mathbf{1}_{\setminus\mathcal{D}})}{ c_{i,\mathcal{G}}(1)}$.

\textit{Proof:} To analyze the asymptotic stability of the $N - D$ dimensional system of differential equations (\ref{eq:differential}) at the point $\mathbf{p}^* = \mathbf{1}_{\setminus\mathcal{D}}$ we apply the Hartman-Grobman theorem and study the corresponding linear system \cite{arrowsmith1990introduction}. 
The linearization asserts that a steady state of the system of differential equations will be locally asymptotically stable when all eigenvalues of its Jacobian $\mathbf{J^{\setminus\mathcal{D}}}(\mathbf{p}^*)$ have negative real parts. To find when this is satisfied, we define $\mathbf{J}^{\setminus\mathcal{D}}_i (\mathbf{p}) = \dot{\mathrm{p}_i}$. Then, the $ij$-th entry of $\mathbf{J}^{\setminus\mathcal{D}}(\mathbf{\hat{p}})$, where $i,j \not\in \mathcal{D}$, at an arbitrary point $\mathbf{\hat{p}}$ of this system can be calculated as
\begin{widetext}
\begin{equation*} 
J^{\setminus\mathcal{D}}_{ij} =  \pdv{\mathbf{J}_i}{\mathrm{p}_j} (\mathbf{\hat{p}}) = 
\begin{cases}
\frac{\dv[2]{f_i\left( f_i^{-1} ( \mathrm{\hat{p}}_i)\right)}{Y} }{\dv{f_i\left( f_i^{-1} ( \mathrm{\hat{p}}_i)\right)}{Y} } \left[ b_{i,\mathcal{G}}\left( \mathbf{\hat{p}} \right) - c_{i,\mathcal{G}} \left( \mathrm{\hat{p}}_i \right) \right] + \dv{\left( f_i^{-1}f_i( \mathrm{\hat{p}}_i)\right)}{Y} \left[ \pdv{b_{i,\mathcal{G}} (\mathbf{\hat{p}})}{\mathrm{p}_i}   + \dv{c_{i,\mathcal{G}}(\mathrm{\hat{p}}_i)}{\mathrm{p}_i} )  \right] ,& \text{if}\ i=j, \\\
\dv{f_i\left( f_i^{-1} ( \mathrm{\hat{p}}_i)\right) }{Y} \pdv{b_{i,\mathcal{G}} (\mathbf{\hat{p}})}{\mathrm{p}_j} ,  & \text{otherwise}.
\end{cases}
\end{equation*}
\end{widetext}

Let $\mathbf{p}^* \to \mathbf{1}_{\setminus\mathcal{D}}$. Since the update rule $f_i$ is monotonically increasing with $\lim_{\mathrm{Y}\to \infty} f_i(\mathrm{Y}) = 1$, we have that $ \lim_{\mathrm{p}_i\to 1} \dv{f_i\left(f_i^{-1} ( \mathrm{\hat{p}}_i)\right)}{Y}  = 0$. Therefore,
\begin{widetext}
\begin{equation} 
\lim_{\mathbf{p}^* \to \mathbf{1}_{\setminus\mathcal{D}}} J^{\setminus\mathcal{D}}_{ij} =  
\begin{cases}
\lim_{\mathbf{p}^* \to \mathbf{1}_{\setminus\mathcal{D}}} \frac{\dv[2]{f_i\left( f_i^{-1} ( \mathrm{p}^*_i)\right)}{Y}}{\dv{f_i\left( f_i^{-1} ( \mathrm{p}^*_i)\right)}{Y} } \left[ b_{i,\mathcal{G}}\left( \mathbf{p}^* \right) - c_{i,\mathcal{G}} \left( \mathrm{p}^*_i \right) \right] ,& \text{if}\ i=j, \\\
0,  & \text{otherwise}.
\end{cases}
\label{eq:jacobian}
\end{equation}
\end{widetext}
This expression simplifies the analysis because we are left with a diagonal matrix, and it is known that in that case the diagonal entries correspond to the eigenvalues of the matrix. Next, by applying the properties of $f_i$ defined with the monotonic transformation $h_i$, i.e. equation (\ref{eq:update_chain_rule}), we find that
\begin{widetext}
\begin{align}
\lim_{\mathbf{p}^* \to \mathbf{1}_{\setminus\mathcal{D}}} J^{\setminus\mathcal{D}}_{ii} &= - \dv{h \left( f^{-1} ( 1 )\right) }{\mathrm{Y}} \left[ b_{i,\mathcal{G}}\left( \mathbf{1}_{\setminus\mathcal{D}} \right) - c_{i,\mathcal{G}} \left( 1 \right) \right].
\label{eq:eigenvalues}
\end{align}
\end{widetext}
Since $\dv{h_i\left( f_i^{-1} ( 1 )\right)}{\mathrm{Y}} = 1$, this expression will be negative if and only if $b_{i,\mathcal{G}} (\mathbf{1}_{\setminus\mathcal{D}}) - c_{i,\mathcal{G}}(1) > 0$, i.e. $v^{\setminus\mathcal{D}}_{i,\mathcal{G}} > 1$. Moreover, because for asymptotic stability (\ref{eq:eigenvalues}) needs to hold for all $i \not\in \mathcal{D}$, we get expression (\ref{eq:full_coop_threshold}). \hfill $\blacksquare$

%


\begin{thebibliography}{50}%
\makeatletter
\providecommand \@ifxundefined [1]{%
 \@ifx{#1\undefined}
}%
\providecommand \@ifnum [1]{%
 \ifnum #1\expandafter \@firstoftwo
 \else \expandafter \@secondoftwo
 \fi
}%
\providecommand \@ifx [1]{%
 \ifx #1\expandafter \@firstoftwo
 \else \expandafter \@secondoftwo
 \fi
}%
\providecommand \natexlab [1]{#1}%
\providecommand \enquote  [1]{``#1''}%
\providecommand \bibnamefont  [1]{#1}%
\providecommand \bibfnamefont [1]{#1}%
\providecommand \citenamefont [1]{#1}%
\providecommand \href@noop [0]{\@secondoftwo}%
\providecommand \href [0]{\begingroup \@sanitize@url \@href}%
\providecommand \@href[1]{\@@startlink{#1}\@@href}%
\providecommand \@@href[1]{\endgroup#1\@@endlink}%
\providecommand \@sanitize@url [0]{\catcode `\\12\catcode `\$12\catcode
  `\&12\catcode `\#12\catcode `\^12\catcode `\_12\catcode `\%12\relax}%
\providecommand \@@startlink[1]{}%
\providecommand \@@endlink[0]{}%
\providecommand \url  [0]{\begingroup\@sanitize@url \@url }%
\providecommand \@url [1]{\endgroup\@href {#1}{\urlprefix }}%
\providecommand \urlprefix  [0]{URL }%
\providecommand \Eprint [0]{\href }%
\providecommand \doibase [0]{http://dx.doi.org/}%
\providecommand \selectlanguage [0]{\@gobble}%
\providecommand \bibinfo  [0]{\@secondoftwo}%
\providecommand \bibfield  [0]{\@secondoftwo}%
\providecommand \translation [1]{[#1]}%
\providecommand \BibitemOpen [0]{}%
\providecommand \bibitemStop [0]{}%
\providecommand \bibitemNoStop [0]{.\EOS\space}%
\providecommand \EOS [0]{\spacefactor3000\relax}%
\providecommand \BibitemShut  [1]{\csname bibitem#1\endcsname}%
\let\auto@bib@innerbib\@empty
\bibitem [{\citenamefont {Capraro}(2013)}]{capraro2013model}%
  \BibitemOpen
  \bibfield  {author} {\bibinfo {author} {\bibfnamefont {V.}~\bibnamefont
  {Capraro}},\ }\href@noop {} {\bibfield  {journal} {\bibinfo  {journal} {PLoS
  One}\ }\textbf {\bibinfo {volume} {8}},\ \bibinfo {pages} {e72427} (\bibinfo
  {year} {2013})}\BibitemShut {NoStop}%
\bibitem [{\citenamefont {Wang}\ \emph {et~al.}(2014)\citenamefont {Wang},
  \citenamefont {Szolnoki},\ and\ \citenamefont {Perc}}]{wang2014different}%
  \BibitemOpen
  \bibfield  {author} {\bibinfo {author} {\bibfnamefont {Z.}~\bibnamefont
  {Wang}}, \bibinfo {author} {\bibfnamefont {A.}~\bibnamefont {Szolnoki}}, \
  and\ \bibinfo {author} {\bibfnamefont {M.}~\bibnamefont {Perc}},\ }\href@noop
  {} {\bibfield  {journal} {\bibinfo  {journal} {Physical Review E}\ }\textbf
  {\bibinfo {volume} {90}},\ \bibinfo {pages} {032813} (\bibinfo {year}
  {2014})}\BibitemShut {NoStop}%
\bibitem [{\citenamefont {Szab{\'o}}\ and\ \citenamefont
  {Bunth}(2018)}]{szabo2018social}%
  \BibitemOpen
  \bibfield  {author} {\bibinfo {author} {\bibfnamefont {G.}~\bibnamefont
  {Szab{\'o}}}\ and\ \bibinfo {author} {\bibfnamefont {G.}~\bibnamefont
  {Bunth}},\ }\href@noop {} {\bibfield  {journal} {\bibinfo  {journal}
  {Physical Review E}\ }\textbf {\bibinfo {volume} {97}},\ \bibinfo {pages}
  {012305} (\bibinfo {year} {2018})}\BibitemShut {NoStop}%
\bibitem [{\citenamefont {Darwin}(1888)}]{darwin1888descent}%
  \BibitemOpen
  \bibfield  {author} {\bibinfo {author} {\bibfnamefont {C.}~\bibnamefont
  {Darwin}},\ }\href@noop {} {\emph {\bibinfo {title} {The descent of man and
  selection in relation to sex}}},\ Vol.~\bibinfo {volume} {1}\ (\bibinfo
  {publisher} {Murray},\ \bibinfo {year} {1888})\BibitemShut {NoStop}%
\bibitem [{\citenamefont {Smith}\ and\ \citenamefont
  {Price}(1973)}]{Smith-1973}%
  \BibitemOpen
  \bibfield  {author} {\bibinfo {author} {\bibfnamefont {J.~M.}\ \bibnamefont
  {Smith}}\ and\ \bibinfo {author} {\bibfnamefont {G.}~\bibnamefont {Price}},\
  }\href@noop {} {\bibfield  {journal} {\bibinfo  {journal} {Nature}\ }\textbf
  {\bibinfo {volume} {246}},\ \bibinfo {pages} {15} (\bibinfo {year}
  {1973})}\BibitemShut {NoStop}%
\bibitem [{\citenamefont {Pennisi}(2005)}]{Pennisi-2005}%
  \BibitemOpen
  \bibfield  {author} {\bibinfo {author} {\bibfnamefont {E.}~\bibnamefont
  {Pennisi}},\ }\href {\doibase 10.1126/science.309.5731.93} {\ \textbf
  {\bibinfo {volume} {309}},\ \bibinfo {pages} {93} (\bibinfo {year}
  {2005})}\BibitemShut {NoStop}%
\bibitem [{\citenamefont {Trivers}(1971)}]{Trivers-1971}%
  \BibitemOpen
  \bibfield  {author} {\bibinfo {author} {\bibfnamefont {R.~L.}\ \bibnamefont
  {Trivers}},\ }\href@noop {} {\bibfield  {journal} {\bibinfo  {journal}
  {Quart. Rev. Biol.}\ ,\ \bibinfo {pages} {35}} (\bibinfo {year}
  {1971})}\BibitemShut {NoStop}%
\bibitem [{\citenamefont {Nowak}\ and\ \citenamefont
  {Sigmund}(2005)}]{nowak2005evolution}%
  \BibitemOpen
  \bibfield  {author} {\bibinfo {author} {\bibfnamefont {M.~A.}\ \bibnamefont
  {Nowak}}\ and\ \bibinfo {author} {\bibfnamefont {K.}~\bibnamefont
  {Sigmund}},\ }\href@noop {} {\bibfield  {journal} {\bibinfo  {journal}
  {Nature}\ ,\ \bibinfo {pages} {1291}} (\bibinfo {year} {2005})}\BibitemShut
  {NoStop}%
\bibitem [{\citenamefont {Axelrod}\ and\ \citenamefont
  {Hamilton}(1981)}]{Axelrod-1981}%
  \BibitemOpen
  \bibfield  {author} {\bibinfo {author} {\bibfnamefont {R.}~\bibnamefont
  {Axelrod}}\ and\ \bibinfo {author} {\bibfnamefont {W.~D.}\ \bibnamefont
  {Hamilton}},\ }\href@noop {} {\bibfield  {journal} {\bibinfo  {journal}
  {Science}\ }\textbf {\bibinfo {volume} {211}},\ \bibinfo {pages} {1390}
  (\bibinfo {year} {1981})}\BibitemShut {NoStop}%
\bibitem [{\citenamefont {Van~Segbroeck}\ \emph {et~al.}(2012)\citenamefont
  {Van~Segbroeck}, \citenamefont {Pacheco}, \citenamefont {Lenaerts},\ and\
  \citenamefont {Santos}}]{van2012emergence}%
  \BibitemOpen
  \bibfield  {author} {\bibinfo {author} {\bibfnamefont {S.}~\bibnamefont
  {Van~Segbroeck}}, \bibinfo {author} {\bibfnamefont {J.~M.}\ \bibnamefont
  {Pacheco}}, \bibinfo {author} {\bibfnamefont {T.}~\bibnamefont {Lenaerts}}, \
  and\ \bibinfo {author} {\bibfnamefont {F.~C.}\ \bibnamefont {Santos}},\
  }\href@noop {} {\bibfield  {journal} {\bibinfo  {journal} {Physical review
  letters}\ }\textbf {\bibinfo {volume} {108}},\ \bibinfo {pages} {158104}
  (\bibinfo {year} {2012})}\BibitemShut {NoStop}%
\bibitem [{\citenamefont {Hauert}\ \emph {et~al.}(2006)\citenamefont {Hauert},
  \citenamefont {Michor}, \citenamefont {Nowak},\ and\ \citenamefont
  {Doebeli}}]{hauert2006synergy}%
  \BibitemOpen
  \bibfield  {author} {\bibinfo {author} {\bibfnamefont {C.}~\bibnamefont
  {Hauert}}, \bibinfo {author} {\bibfnamefont {F.}~\bibnamefont {Michor}},
  \bibinfo {author} {\bibfnamefont {M.~A.}\ \bibnamefont {Nowak}}, \ and\
  \bibinfo {author} {\bibfnamefont {M.}~\bibnamefont {Doebeli}},\ }\href@noop
  {} {\bibfield  {journal} {\bibinfo  {journal} {Journal of theoretical
  biology}\ }\textbf {\bibinfo {volume} {239}},\ \bibinfo {pages} {195}
  (\bibinfo {year} {2006})}\BibitemShut {NoStop}%
\bibitem [{\citenamefont {Killingback}\ and\ \citenamefont
  {Doebeli}(2002)}]{killingback2002continuous}%
  \BibitemOpen
  \bibfield  {author} {\bibinfo {author} {\bibfnamefont {T.}~\bibnamefont
  {Killingback}}\ and\ \bibinfo {author} {\bibfnamefont {M.}~\bibnamefont
  {Doebeli}},\ }\href@noop {} {\bibfield  {journal} {\bibinfo  {journal} {The
  American Naturalist}\ }\textbf {\bibinfo {volume} {160}},\ \bibinfo {pages}
  {421} (\bibinfo {year} {2002})}\BibitemShut {NoStop}%
\bibitem [{\citenamefont {Yoeli}\ \emph {et~al.}(2013)\citenamefont {Yoeli},
  \citenamefont {Hoffman}, \citenamefont {Rand},\ and\ \citenamefont
  {Nowak}}]{yoeli2013powering}%
  \BibitemOpen
  \bibfield  {author} {\bibinfo {author} {\bibfnamefont {E.}~\bibnamefont
  {Yoeli}}, \bibinfo {author} {\bibfnamefont {M.}~\bibnamefont {Hoffman}},
  \bibinfo {author} {\bibfnamefont {D.~G.}\ \bibnamefont {Rand}}, \ and\
  \bibinfo {author} {\bibfnamefont {M.~A.}\ \bibnamefont {Nowak}},\ }\href@noop
  {} {\bibfield  {journal} {\bibinfo  {journal} {Proceedings of the National
  Academy of Sciences}\ }\textbf {\bibinfo {volume} {110}},\ \bibinfo {pages}
  {10424} (\bibinfo {year} {2013})}\BibitemShut {NoStop}%
\bibitem [{\citenamefont {Hilbe}\ \emph {et~al.}(2014)\citenamefont {Hilbe},
  \citenamefont {Wu}, \citenamefont {Traulsen},\ and\ \citenamefont
  {Nowak}}]{hilbe2014cooperation}%
  \BibitemOpen
  \bibfield  {author} {\bibinfo {author} {\bibfnamefont {C.}~\bibnamefont
  {Hilbe}}, \bibinfo {author} {\bibfnamefont {B.}~\bibnamefont {Wu}}, \bibinfo
  {author} {\bibfnamefont {A.}~\bibnamefont {Traulsen}}, \ and\ \bibinfo
  {author} {\bibfnamefont {M.~A.}\ \bibnamefont {Nowak}},\ }\href@noop {}
  {\bibfield  {journal} {\bibinfo  {journal} {Proceedings of the National
  Academy of Sciences}\ }\textbf {\bibinfo {volume} {111}},\ \bibinfo {pages}
  {16425} (\bibinfo {year} {2014})}\BibitemShut {NoStop}%
\bibitem [{\citenamefont {Hilbe}\ \emph {et~al.}(2015)\citenamefont {Hilbe},
  \citenamefont {Wu}, \citenamefont {Traulsen},\ and\ \citenamefont
  {Nowak}}]{hilbe2015evolutionary}%
  \BibitemOpen
  \bibfield  {author} {\bibinfo {author} {\bibfnamefont {C.}~\bibnamefont
  {Hilbe}}, \bibinfo {author} {\bibfnamefont {B.}~\bibnamefont {Wu}}, \bibinfo
  {author} {\bibfnamefont {A.}~\bibnamefont {Traulsen}}, \ and\ \bibinfo
  {author} {\bibfnamefont {M.~A.}\ \bibnamefont {Nowak}},\ }\href@noop {}
  {\bibfield  {journal} {\bibinfo  {journal} {Journal of theoretical biology}\
  }\textbf {\bibinfo {volume} {374}},\ \bibinfo {pages} {115} (\bibinfo {year}
  {2015})}\BibitemShut {NoStop}%
\bibitem [{\citenamefont {Pan}\ \emph {et~al.}(2015)\citenamefont {Pan},
  \citenamefont {Hao}, \citenamefont {Rong},\ and\ \citenamefont
  {Zhou}}]{pan2015zero}%
  \BibitemOpen
  \bibfield  {author} {\bibinfo {author} {\bibfnamefont {L.}~\bibnamefont
  {Pan}}, \bibinfo {author} {\bibfnamefont {D.}~\bibnamefont {Hao}}, \bibinfo
  {author} {\bibfnamefont {Z.}~\bibnamefont {Rong}}, \ and\ \bibinfo {author}
  {\bibfnamefont {T.}~\bibnamefont {Zhou}},\ }\href@noop {} {\bibfield
  {journal} {\bibinfo  {journal} {Scientific reports}\ }\textbf {\bibinfo
  {volume} {5}} (\bibinfo {year} {2015})}\BibitemShut {NoStop}%
\bibitem [{\citenamefont {van Doorn}\ and\ \citenamefont
  {Taborsky}(2012)}]{vanDoorn-2012}%
  \BibitemOpen
  \bibfield  {author} {\bibinfo {author} {\bibfnamefont {G.~S.}\ \bibnamefont
  {van Doorn}}\ and\ \bibinfo {author} {\bibfnamefont {M.}~\bibnamefont
  {Taborsky}},\ }\href {\doibase 10.1111/j.1558-5646.2011.01479.x} {\bibfield
  {journal} {\bibinfo  {journal} {Evolution}\ }\textbf {\bibinfo {volume}
  {66}},\ \bibinfo {pages} {651} (\bibinfo {year} {2012})}\BibitemShut
  {NoStop}%
\bibitem [{\citenamefont {Taborsky}\ \emph {et~al.}(2016)\citenamefont
  {Taborsky}, \citenamefont {Frommen},\ and\ \citenamefont
  {Riehl}}]{Taborsky-2016}%
  \BibitemOpen
  \bibfield  {author} {\bibinfo {author} {\bibfnamefont {M.}~\bibnamefont
  {Taborsky}}, \bibinfo {author} {\bibfnamefont {J.~G.}\ \bibnamefont
  {Frommen}}, \ and\ \bibinfo {author} {\bibfnamefont {C.}~\bibnamefont
  {Riehl}},\ }\href
  {http://rstb.royalsocietypublishing.org/content/371/1687/20150084} {\bibfield
   {journal} {\bibinfo  {journal} {Philos. Trans. Roy. Soc. B.}\ }\textbf
  {\bibinfo {volume} {371}} (\bibinfo {year} {2016})}\BibitemShut {NoStop}%
\bibitem [{\citenamefont {Pfeiffer}\ \emph {et~al.}(2005)\citenamefont
  {Pfeiffer}, \citenamefont {Rutte}, \citenamefont {Killingback}, \citenamefont
  {Taborsky},\ and\ \citenamefont {Bonhoeffer}}]{pfeiffer2005evolution}%
  \BibitemOpen
  \bibfield  {author} {\bibinfo {author} {\bibfnamefont {T.}~\bibnamefont
  {Pfeiffer}}, \bibinfo {author} {\bibfnamefont {C.}~\bibnamefont {Rutte}},
  \bibinfo {author} {\bibfnamefont {T.}~\bibnamefont {Killingback}}, \bibinfo
  {author} {\bibfnamefont {M.}~\bibnamefont {Taborsky}}, \ and\ \bibinfo
  {author} {\bibfnamefont {S.}~\bibnamefont {Bonhoeffer}},\ }\href@noop {}
  {\bibfield  {journal} {\bibinfo  {journal} {Proceedings of the Royal Society
  of London B: Biological Sciences}\ }\textbf {\bibinfo {volume} {272}},\
  \bibinfo {pages} {1115} (\bibinfo {year} {2005})}\BibitemShut {NoStop}%
\bibitem [{\citenamefont {Boyd}\ and\ \citenamefont
  {Richerson}(1989)}]{Boyd-1989}%
  \BibitemOpen
  \bibfield  {author} {\bibinfo {author} {\bibfnamefont {R.}~\bibnamefont
  {Boyd}}\ and\ \bibinfo {author} {\bibfnamefont {P.}~\bibnamefont
  {Richerson}},\ }\href {\doibase 10.1016/0378-8733(89)90003-8} {\bibfield
  {journal} {\bibinfo  {journal} {Soc. Networks}\ }\textbf {\bibinfo {volume}
  {11}},\ \bibinfo {pages} {213} (\bibinfo {year} {1989})}\BibitemShut
  {NoStop}%
\bibitem [{\citenamefont {Nowak}\ and\ \citenamefont
  {Roch}(2007)}]{Nowak-2005}%
  \BibitemOpen
  \bibfield  {author} {\bibinfo {author} {\bibfnamefont {M.~A.}\ \bibnamefont
  {Nowak}}\ and\ \bibinfo {author} {\bibfnamefont {S.}~\bibnamefont {Roch}},\
  }\href {\doibase 10.1098/rspb.2006.0125} {\bibfield  {journal} {\bibinfo
  {journal} {Philos. Trans. Roy. Soc. B.}\ }\textbf {\bibinfo {volume} {274}},\
  \bibinfo {pages} {605} (\bibinfo {year} {2007})}\BibitemShut {NoStop}%
\bibitem [{\citenamefont {Rutte}\ and\ \citenamefont
  {Taborsky}(2007)}]{Rutte-2007}%
  \BibitemOpen
  \bibfield  {author} {\bibinfo {author} {\bibfnamefont {C.}~\bibnamefont
  {Rutte}}\ and\ \bibinfo {author} {\bibfnamefont {M.}~\bibnamefont
  {Taborsky}},\ }\href@noop {} {\bibfield  {journal} {\bibinfo  {journal} {PLoS
  Biol}\ }\textbf {\bibinfo {volume} {5}},\ \bibinfo {pages} {e196} (\bibinfo
  {year} {2007})}\BibitemShut {NoStop}%
\bibitem [{\citenamefont {Bartlett}\ and\ \citenamefont
  {DeSteno}(2006)}]{Bartlett-2006}%
  \BibitemOpen
  \bibfield  {author} {\bibinfo {author} {\bibfnamefont {M.~Y.}\ \bibnamefont
  {Bartlett}}\ and\ \bibinfo {author} {\bibfnamefont {D.}~\bibnamefont
  {DeSteno}},\ }\href@noop {} {\bibfield  {journal} {\bibinfo  {journal}
  {Psychol. Sci.}\ }\textbf {\bibinfo {volume} {17}},\ \bibinfo {pages} {319}
  (\bibinfo {year} {2006})}\BibitemShut {NoStop}%
\bibitem [{\citenamefont {Isen}(1987)}]{isen1987positive}%
  \BibitemOpen
  \bibfield  {author} {\bibinfo {author} {\bibfnamefont {A.~M.}\ \bibnamefont
  {Isen}},\ }in\ \href@noop {} {\emph {\bibinfo {booktitle} {Advances in
  experimental social psychology}}},\ Vol.~\bibinfo {volume} {20}\ (\bibinfo
  {publisher} {Elsevier},\ \bibinfo {year} {1987})\ pp.\ \bibinfo {pages}
  {203--253}\BibitemShut {NoStop}%
\bibitem [{\citenamefont {Leimgruber}\ \emph {et~al.}(2014)\citenamefont
  {Leimgruber}, \citenamefont {Ward}, \citenamefont {Widness}, \citenamefont
  {Norton}, \citenamefont {Olson}, \citenamefont {Gray},\ and\ \citenamefont
  {Santos}}]{Leimgruber-2014}%
  \BibitemOpen
  \bibfield  {author} {\bibinfo {author} {\bibfnamefont {K.~L.}\ \bibnamefont
  {Leimgruber}}, \bibinfo {author} {\bibfnamefont {A.~F.}\ \bibnamefont
  {Ward}}, \bibinfo {author} {\bibfnamefont {J.}~\bibnamefont {Widness}},
  \bibinfo {author} {\bibfnamefont {M.~I.}\ \bibnamefont {Norton}}, \bibinfo
  {author} {\bibfnamefont {K.~R.}\ \bibnamefont {Olson}}, \bibinfo {author}
  {\bibfnamefont {K.}~\bibnamefont {Gray}}, \ and\ \bibinfo {author}
  {\bibfnamefont {L.~R.}\ \bibnamefont {Santos}},\ }\href@noop {} {\bibfield
  {journal} {\bibinfo  {journal} {PloS one}\ }\textbf {\bibinfo {volume} {9}},\
  \bibinfo {pages} {e87035} (\bibinfo {year} {2014})}\BibitemShut {NoStop}%
\bibitem [{\citenamefont {Gfrerer}\ and\ \citenamefont
  {Taborsky}(2017)}]{Gfrerer-2017}%
  \BibitemOpen
  \bibfield  {author} {\bibinfo {author} {\bibfnamefont {N.}~\bibnamefont
  {Gfrerer}}\ and\ \bibinfo {author} {\bibfnamefont {M.}~\bibnamefont
  {Taborsky}},\ }\href@noop {} {\bibfield  {journal} {\bibinfo  {journal} {Sci.
  Rep.}\ }\textbf {\bibinfo {volume} {7}} (\bibinfo {year} {2017})}\BibitemShut
  {NoStop}%
\bibitem [{\citenamefont {Stanca}(2009)}]{Stanca-2009}%
  \BibitemOpen
  \bibfield  {author} {\bibinfo {author} {\bibfnamefont {L.}~\bibnamefont
  {Stanca}},\ }\href@noop {} {\bibfield  {journal} {\bibinfo  {journal} {J.
  Econ. Psychol.}\ }\textbf {\bibinfo {volume} {30}},\ \bibinfo {pages} {190}
  (\bibinfo {year} {2009})}\BibitemShut {NoStop}%
\bibitem [{\citenamefont {Barta}\ \emph {et~al.}(2011)\citenamefont {Barta},
  \citenamefont {McNamara}, \citenamefont {Husz{\'a}r},\ and\ \citenamefont
  {Taborsky}}]{Barta-2011}%
  \BibitemOpen
  \bibfield  {author} {\bibinfo {author} {\bibfnamefont {Z.}~\bibnamefont
  {Barta}}, \bibinfo {author} {\bibfnamefont {J.~M.}\ \bibnamefont {McNamara}},
  \bibinfo {author} {\bibfnamefont {D.~B.}\ \bibnamefont {Husz{\'a}r}}, \ and\
  \bibinfo {author} {\bibfnamefont {M.}~\bibnamefont {Taborsky}},\ }\href@noop
  {} {\bibfield  {journal} {\bibinfo  {journal} {Philos. Trans. Roy. Soc. B.}\
  }\textbf {\bibinfo {volume} {278}},\ \bibinfo {pages} {843} (\bibinfo {year}
  {2011})}\BibitemShut {NoStop}%
\bibitem [{\citenamefont {Utkovski}\ \emph {et~al.}(2017)\citenamefont
  {Utkovski}, \citenamefont {Stojkoski}, \citenamefont {Basnarkov},\ and\
  \citenamefont {Kocarev}}]{Utkovski-2017}%
  \BibitemOpen
  \bibfield  {author} {\bibinfo {author} {\bibfnamefont {Z.}~\bibnamefont
  {Utkovski}}, \bibinfo {author} {\bibfnamefont {V.}~\bibnamefont {Stojkoski}},
  \bibinfo {author} {\bibfnamefont {L.}~\bibnamefont {Basnarkov}}, \ and\
  \bibinfo {author} {\bibfnamefont {L.}~\bibnamefont {Kocarev}},\ }\href@noop
  {} {\bibfield  {journal} {\bibinfo  {journal} {Physical Review E}\ }\textbf
  {\bibinfo {volume} {96}},\ \bibinfo {pages} {022315} (\bibinfo {year}
  {2017})}\BibitemShut {NoStop}%
\bibitem [{\citenamefont {Perc}\ \emph {et~al.}(2017)\citenamefont {Perc},
  \citenamefont {Jordan}, \citenamefont {Rand}, \citenamefont {Wang},
  \citenamefont {Boccaletti},\ and\ \citenamefont {Szolnoki}}]{Perc-2017}%
  \BibitemOpen
  \bibfield  {author} {\bibinfo {author} {\bibfnamefont {M.}~\bibnamefont
  {Perc}}, \bibinfo {author} {\bibfnamefont {J.~J.}\ \bibnamefont {Jordan}},
  \bibinfo {author} {\bibfnamefont {D.~G.}\ \bibnamefont {Rand}}, \bibinfo
  {author} {\bibfnamefont {Z.}~\bibnamefont {Wang}}, \bibinfo {author}
  {\bibfnamefont {S.}~\bibnamefont {Boccaletti}}, \ and\ \bibinfo {author}
  {\bibfnamefont {A.}~\bibnamefont {Szolnoki}},\ }\href@noop {} {\bibfield
  {journal} {\bibinfo  {journal} {Physics Reports}\ }\textbf {\bibinfo {volume}
  {687}},\ \bibinfo {pages} {1} (\bibinfo {year} {2017})}\BibitemShut {NoStop}%
\bibitem [{\citenamefont {Milinski}\ and\ \citenamefont
  {Rockenbach}(2012)}]{milinski2012interaction}%
  \BibitemOpen
  \bibfield  {author} {\bibinfo {author} {\bibfnamefont {M.}~\bibnamefont
  {Milinski}}\ and\ \bibinfo {author} {\bibfnamefont {B.}~\bibnamefont
  {Rockenbach}},\ }\href@noop {} {\bibfield  {journal} {\bibinfo  {journal}
  {Journal of Theoretical Biology}\ }\textbf {\bibinfo {volume} {299}},\
  \bibinfo {pages} {139} (\bibinfo {year} {2012})}\BibitemShut {NoStop}%
\bibitem [{\citenamefont {Motro}(1991)}]{motro1991co}%
  \BibitemOpen
  \bibfield  {author} {\bibinfo {author} {\bibfnamefont {U.}~\bibnamefont
  {Motro}},\ }\href@noop {} {\bibfield  {journal} {\bibinfo  {journal} {Journal
  of Theoretical Biology}\ }\textbf {\bibinfo {volume} {151}},\ \bibinfo
  {pages} {145} (\bibinfo {year} {1991})}\BibitemShut {NoStop}%
\bibitem [{\citenamefont {Kerr}\ \emph {et~al.}(2004)\citenamefont {Kerr},
  \citenamefont {Godfrey-Smith},\ and\ \citenamefont
  {Feldman}}]{kerr2004altruism}%
  \BibitemOpen
  \bibfield  {author} {\bibinfo {author} {\bibfnamefont {B.}~\bibnamefont
  {Kerr}}, \bibinfo {author} {\bibfnamefont {P.}~\bibnamefont {Godfrey-Smith}},
  \ and\ \bibinfo {author} {\bibfnamefont {M.~W.}\ \bibnamefont {Feldman}},\
  }\href@noop {} {\bibfield  {journal} {\bibinfo  {journal} {Trends in ecology
  \& evolution}\ }\textbf {\bibinfo {volume} {19}},\ \bibinfo {pages} {135}
  (\bibinfo {year} {2004})}\BibitemShut {NoStop}%
\bibitem [{\citenamefont {Nowak}(2006)}]{Nowak-2006five}%
  \BibitemOpen
  \bibfield  {author} {\bibinfo {author} {\bibfnamefont {M.~A.}\ \bibnamefont
  {Nowak}},\ }\href@noop {} {\bibfield  {journal} {\bibinfo  {journal}
  {Science}\ }\textbf {\bibinfo {volume} {314}},\ \bibinfo {pages} {1560}
  (\bibinfo {year} {2006})}\BibitemShut {NoStop}%
\bibitem [{\citenamefont {Ostrom}(2015)}]{ostrom2015governing}%
  \BibitemOpen
  \bibfield  {author} {\bibinfo {author} {\bibfnamefont {E.}~\bibnamefont
  {Ostrom}},\ }\href@noop {} {\emph {\bibinfo {title} {Governing the
  commons}}}\ (\bibinfo  {publisher} {Cambridge university press},\ \bibinfo
  {year} {2015})\BibitemShut {NoStop}%
\bibitem [{\citenamefont {Santos}\ and\ \citenamefont
  {Pacheco}(2005)}]{Santos-2005}%
  \BibitemOpen
  \bibfield  {author} {\bibinfo {author} {\bibfnamefont {F.~C.}\ \bibnamefont
  {Santos}}\ and\ \bibinfo {author} {\bibfnamefont {J.~M.}\ \bibnamefont
  {Pacheco}},\ }\href {\doibase 10.1103/PhysRevLett.95.098104} {\bibfield
  {journal} {\bibinfo  {journal} {Phys. Rev. Lett.}\ }\textbf {\bibinfo
  {volume} {95}},\ \bibinfo {pages} {098104} (\bibinfo {year}
  {2005})}\BibitemShut {NoStop}%
\bibitem [{\citenamefont {Doebeli}\ and\ \citenamefont
  {Hauert}(2005)}]{Doebeli-2005}%
  \BibitemOpen
  \bibfield  {author} {\bibinfo {author} {\bibfnamefont {M.}~\bibnamefont
  {Doebeli}}\ and\ \bibinfo {author} {\bibfnamefont {C.}~\bibnamefont
  {Hauert}},\ }\href@noop {} {\bibfield  {journal} {\bibinfo  {journal}
  {Ecology Letters}\ }\textbf {\bibinfo {volume} {8}},\ \bibinfo {pages} {748}
  (\bibinfo {year} {2005})}\BibitemShut {NoStop}%
\bibitem [{\citenamefont {Archetti}\ and\ \citenamefont
  {Scheuring}(2012)}]{archetti2012game}%
  \BibitemOpen
  \bibfield  {author} {\bibinfo {author} {\bibfnamefont {M.}~\bibnamefont
  {Archetti}}\ and\ \bibinfo {author} {\bibfnamefont {I.}~\bibnamefont
  {Scheuring}},\ }\href@noop {} {\bibfield  {journal} {\bibinfo  {journal}
  {Journal of theoretical biology}\ }\textbf {\bibinfo {volume} {299}},\
  \bibinfo {pages} {9} (\bibinfo {year} {2012})}\BibitemShut {NoStop}%
\bibitem [{\citenamefont {Battiston}\ \emph {et~al.}(2017)\citenamefont
  {Battiston}, \citenamefont {Perc},\ and\ \citenamefont
  {Latora}}]{battiston2017determinants}%
  \BibitemOpen
  \bibfield  {author} {\bibinfo {author} {\bibfnamefont {F.}~\bibnamefont
  {Battiston}}, \bibinfo {author} {\bibfnamefont {M.}~\bibnamefont {Perc}}, \
  and\ \bibinfo {author} {\bibfnamefont {V.}~\bibnamefont {Latora}},\
  }\href@noop {} {\bibfield  {journal} {\bibinfo  {journal} {New Journal of
  Physics}\ } (\bibinfo {year} {2017})}\BibitemShut {NoStop}%
\bibitem [{\citenamefont {Hilbe}\ \emph {et~al.}(2017)\citenamefont {Hilbe},
  \citenamefont {Martinez-Vaquero}, \citenamefont {Chatterjee},\ and\
  \citenamefont {Nowak}}]{hilbe2017memory}%
  \BibitemOpen
  \bibfield  {author} {\bibinfo {author} {\bibfnamefont {C.}~\bibnamefont
  {Hilbe}}, \bibinfo {author} {\bibfnamefont {L.~A.}\ \bibnamefont
  {Martinez-Vaquero}}, \bibinfo {author} {\bibfnamefont {K.}~\bibnamefont
  {Chatterjee}}, \ and\ \bibinfo {author} {\bibfnamefont {M.~A.}\ \bibnamefont
  {Nowak}},\ }\href@noop {} {\bibfield  {journal} {\bibinfo  {journal}
  {Proceedings of the National Academy of Sciences}\ }\textbf {\bibinfo
  {volume} {114}},\ \bibinfo {pages} {4715} (\bibinfo {year}
  {2017})}\BibitemShut {NoStop}%
\bibitem [{\citenamefont {Santos}\ \emph {et~al.}(2008)\citenamefont {Santos},
  \citenamefont {Santos},\ and\ \citenamefont {Pacheco}}]{santos2008social}%
  \BibitemOpen
  \bibfield  {author} {\bibinfo {author} {\bibfnamefont {F.~C.}\ \bibnamefont
  {Santos}}, \bibinfo {author} {\bibfnamefont {M.~D.}\ \bibnamefont {Santos}},
  \ and\ \bibinfo {author} {\bibfnamefont {J.~M.}\ \bibnamefont {Pacheco}},\
  }\href@noop {} {\bibfield  {journal} {\bibinfo  {journal} {Nature}\ }\textbf
  {\bibinfo {volume} {454}},\ \bibinfo {pages} {213} (\bibinfo {year}
  {2008})}\BibitemShut {NoStop}%
\bibitem [{\citenamefont {Perc}\ \emph {et~al.}(2013)\citenamefont {Perc},
  \citenamefont {G{\'o}mez-Garde{\~n}es}, \citenamefont {Szolnoki},
  \citenamefont {Flor{\'\i}a},\ and\ \citenamefont
  {Moreno}}]{perc2013evolutionary}%
  \BibitemOpen
  \bibfield  {author} {\bibinfo {author} {\bibfnamefont {M.}~\bibnamefont
  {Perc}}, \bibinfo {author} {\bibfnamefont {J.}~\bibnamefont
  {G{\'o}mez-Garde{\~n}es}}, \bibinfo {author} {\bibfnamefont {A.}~\bibnamefont
  {Szolnoki}}, \bibinfo {author} {\bibfnamefont {L.~M.}\ \bibnamefont
  {Flor{\'\i}a}}, \ and\ \bibinfo {author} {\bibfnamefont {Y.}~\bibnamefont
  {Moreno}},\ }\href@noop {} {\bibfield  {journal} {\bibinfo  {journal}
  {Journal of the royal society interface}\ }\textbf {\bibinfo {volume} {10}},\
  \bibinfo {pages} {20120997} (\bibinfo {year} {2013})}\BibitemShut {NoStop}%
\bibitem [{\citenamefont {Baker}\ and\ \citenamefont
  {Bulkley}(2014)}]{baker2014paying}%
  \BibitemOpen
  \bibfield  {author} {\bibinfo {author} {\bibfnamefont {W.~E.}\ \bibnamefont
  {Baker}}\ and\ \bibinfo {author} {\bibfnamefont {N.}~\bibnamefont
  {Bulkley}},\ }\href@noop {} {\bibfield  {journal} {\bibinfo  {journal}
  {Organization science}\ }\textbf {\bibinfo {volume} {25}},\ \bibinfo {pages}
  {1493} (\bibinfo {year} {2014})}\BibitemShut {NoStop}%
\bibitem [{\citenamefont {Nicosia}\ \emph {et~al.}(2017)\citenamefont
  {Nicosia}, \citenamefont {Skardal}, \citenamefont {Arenas},\ and\
  \citenamefont {Latora}}]{nicosia2017collective}%
  \BibitemOpen
  \bibfield  {author} {\bibinfo {author} {\bibfnamefont {V.}~\bibnamefont
  {Nicosia}}, \bibinfo {author} {\bibfnamefont {P.~S.}\ \bibnamefont
  {Skardal}}, \bibinfo {author} {\bibfnamefont {A.}~\bibnamefont {Arenas}}, \
  and\ \bibinfo {author} {\bibfnamefont {V.}~\bibnamefont {Latora}},\
  }\href@noop {} {\bibfield  {journal} {\bibinfo  {journal} {Physical review
  letters}\ }\textbf {\bibinfo {volume} {118}},\ \bibinfo {pages} {138302}
  (\bibinfo {year} {2017})}\BibitemShut {NoStop}%
\bibitem [{\citenamefont {Antonioni}\ and\ \citenamefont
  {Cardillo}(2017)}]{antonioni2017coevolution}%
  \BibitemOpen
  \bibfield  {author} {\bibinfo {author} {\bibfnamefont {A.}~\bibnamefont
  {Antonioni}}\ and\ \bibinfo {author} {\bibfnamefont {A.}~\bibnamefont
  {Cardillo}},\ }\href@noop {} {\bibfield  {journal} {\bibinfo  {journal}
  {Physical review letters}\ }\textbf {\bibinfo {volume} {118}},\ \bibinfo
  {pages} {238301} (\bibinfo {year} {2017})}\BibitemShut {NoStop}%
\bibitem [{\citenamefont {Hardin}(2009)}]{Hardin-2009}%
  \BibitemOpen
  \bibfield  {author} {\bibinfo {author} {\bibfnamefont {G.}~\bibnamefont
  {Hardin}},\ }\href@noop {} {\bibfield  {journal} {\bibinfo  {journal}
  {Journal of Natural Resources Policy Research}\ }\textbf {\bibinfo {volume}
  {1}},\ \bibinfo {pages} {243} (\bibinfo {year} {2009})}\BibitemShut {NoStop}%
\bibitem [{\citenamefont {Brechner}(1977)}]{Brechner-1977}%
  \BibitemOpen
  \bibfield  {author} {\bibinfo {author} {\bibfnamefont {K.~C.}\ \bibnamefont
  {Brechner}},\ }\href@noop {} {\bibfield  {journal} {\bibinfo  {journal}
  {Journal of experimental social psychology}\ }\textbf {\bibinfo {volume}
  {13}},\ \bibinfo {pages} {552} (\bibinfo {year} {1977})}\BibitemShut
  {NoStop}%
\bibitem [{\citenamefont {Sugiarto}\ \emph {et~al.}(2017)\citenamefont
  {Sugiarto}, \citenamefont {Lansing}, \citenamefont {Chung}, \citenamefont
  {Lai}, \citenamefont {Cheong},\ and\ \citenamefont
  {Chew}}]{sugiarto2017social}%
  \BibitemOpen
  \bibfield  {author} {\bibinfo {author} {\bibfnamefont {H.~S.}\ \bibnamefont
  {Sugiarto}}, \bibinfo {author} {\bibfnamefont {J.~S.}\ \bibnamefont
  {Lansing}}, \bibinfo {author} {\bibfnamefont {N.~N.}\ \bibnamefont {Chung}},
  \bibinfo {author} {\bibfnamefont {C.}~\bibnamefont {Lai}}, \bibinfo {author}
  {\bibfnamefont {S.~A.}\ \bibnamefont {Cheong}}, \ and\ \bibinfo {author}
  {\bibfnamefont {L.~Y.}\ \bibnamefont {Chew}},\ }\href@noop {} {\bibfield
  {journal} {\bibinfo  {journal} {Physical Review Letters}\ }\textbf {\bibinfo
  {volume} {118}},\ \bibinfo {pages} {208301} (\bibinfo {year}
  {2017})}\BibitemShut {NoStop}%
\bibitem [{\citenamefont {De~la Fuente}(2000)}]{de2000mathematical}%
  \BibitemOpen
  \bibfield  {author} {\bibinfo {author} {\bibfnamefont {A.}~\bibnamefont
  {De~la Fuente}},\ }\href@noop {} {\emph {\bibinfo {title} {Mathematical
  methods and models for economists}}}\ (\bibinfo  {publisher} {Cambridge
  University Press},\ \bibinfo {year} {2000})\BibitemShut {NoStop}%
\bibitem [{\citenamefont {Arrowsmith}\ and\ \citenamefont
  {Place}(1990)}]{arrowsmith1990introduction}%
  \BibitemOpen
  \bibfield  {author} {\bibinfo {author} {\bibfnamefont {D.~K.}\ \bibnamefont
  {Arrowsmith}}\ and\ \bibinfo {author} {\bibfnamefont {C.~M.}\ \bibnamefont
  {Place}},\ }\href@noop {} {\emph {\bibinfo {title} {An introduction to
  dynamical systems}}}\ (\bibinfo  {publisher} {Cambridge university press},\
  \bibinfo {year} {1990})\BibitemShut {NoStop}%
\end{thebibliography}
\end{document}